\documentclass{article} 

\usepackage{mathtools}

\usepackage[a4paper,top=3cm,bottom=2cm,left=3cm,right=3cm,marginparwidth=1.75cm]{geometry}

\usepackage{amssymb}
\usepackage{graphicx}
\usepackage{subcaption}
\usepackage{floatrow}
\usepackage{natbib}

\title{A versatile scanning method for volumetric measurements of velocity and density fields}
\author{J. L. Partridge, A. Lefauve, Stuart B. Dalziel}
\date{}

\begin{document}

\captionsetup[subfigure]{labelfont=it}

\maketitle

\begin{abstract}
Understanding turbulence in a stratified environment requires a detailed picture of both the velocity field and the density field. Experimentally, this represents a significant measurement challenge, especially when full three-dimensional data is needed to accurately characterise the turbulent fields. This paper presents a new approach to obtaining such data through well-resolved, near-instantaneous volume-spanning measurements. This is accomplished by rapidly scanning the volume with a light sheet so that, at each scan location, planar two-dimensional measurements of three velocity components via stereo particle image velocimetry (PIV) and simultaneous density information via planar laser induced fluorescence (PLIF) can be made. Ultimately, this rapid scanning technique permits the measurement of all three components of velocity in a volume as well as simultaneously capturing the three-dimensional density field. The scanning of the light sheet is accomplished by mounting the optics producing the light sheet on a linear traverse that is capable of rapidly scanning the volume in a continuous manner. The key and novel aspect enabling high scan rates is the addition of two mirrors on galvanometers to make small adjustments to the position of the light sheet and ensure a precise overlap between pairs of image frames. This new technique means the light sheet does not have to be excessively thick, the scanning speed too slow, or that inappropriately small interframe times have to be used, while ensuring overlapping particle patterns between pairs of images that are required by the PIV algorithm. The technique is illustrated with some preliminary results from the buoyancy-driven exchange flow through an inclined duct connecting two reservoirs containing different density fluids.
\end{abstract}

\section{Introduction}
Buoyancy forces (due to density variations) play an important role in many flows. Such flows are typically turbulent, exhibiting a range of scales and chaotic motion characterised by a large Reynolds number $Re=\frac{UL}{\nu}>>1$. Here, $U$ is a typical velocity scale, $L$ is a typical length scale and $\nu$ is the kinematic viscosity. Despite advances in experimental measurement techniques, capturing near-instantaneous, simultaneous volumetric density and velocity fields is challenging. Nevertheless, having both density and velocity information is desirable in understanding stratified turbulence due to the increasing interest in analysing coherent structures rather than just statistical quantities. Studying coherent structures has revealed new insight in unstratified (for example, \citet{kerswell2007recurrence} and \citet{kawahara2012significance}) and, recently, stratified \citep{2017arXiv170105406L} turbulent flows. Moreover, there is still a need for experimental measurements of realisable flows to complement direct numerical simulations (DNS). This is due to the complexity of modelling certain physical boundary conditions and domains, as well as, in stratified flows, the non-trivial dependence on the Schmidt number $Sc=\frac{\nu}{\kappa}$, where $\kappa$ is the molecular diffusivity \citep{zhou2017diapycnal}. High $Sc$ number flows (e.g. $Sc=O(10^3)$ in the salinity-stratified ocean) require fine resolution DNS to capture the Bachelor scale 
\begin{equation}
\eta_b = \frac{\eta_k}{Sc^{1/2}}, 
\end{equation}
where 
\begin{equation}
\eta_k=\left(\frac{\nu^3}{\epsilon}\right)^{1/4},
\end{equation}
is the Kolmogorov scale with $\epsilon$ the dissipation rate of turbulent kinetic energy. Given that $\eta_b$ is $Sc^{1/2}$ smaller than $\eta_k$, it is clear that, to resolve the Batchelor scale, high resolution DNS is needed to capture high $Sc$ number flows \citep{schumacher2005very}. Currently, this makes high $Sc$ DNS {prohibitive,} even at the relatively low $Re$ that are easily achieved using salt-stratified (high $Sc$) laboratory experiments. To this end, we present a new method of experimentally capturing simultaneous volumetric density and velocity fields, capable of capturing both high $Re$ and high $Sc$ laboratory flows. 

The layout of the paper is as follows. In \S\ref{review} we review some of the current methodologies to experimentally determine velocity and density fields before describing our new approach in \S\ref{method}. Our experimental set-up, including the specific hardware used to perform our new measurements, is detailed in \S\ref{expSetup} before summarising our results in \S\ref{results}. Possible extensions of the methodology presented in this paper are detailed in \S\ref{discuss}. Finally, conclusions are given in \S\ref{conc}.   

\section{Review of Approaches}\label{review}

In this section we give a brief review of the methodologies commonly used to experimentally measure velocity and density fields. 
\subsection{Planar Velocity Measurements}\label{standardPIV}

Experimentally, measuring the velocity of a flow is commonly achieved using particle tracking velocimetry (PTV) or particle image velocimetry (PIV). In liquid flows, small diameter ($O(10)$ $\mu$m) particles are seeded within the working fluid. The particle diameter $d_p$ and density $\rho_p$ are chosen such that the Stokes number is small $St \equiv \frac{\tau_{p}}{\tau_{\eta}} = \frac{\rho_{p}{d_p}^2}{18\rho\eta_k^2}<< 1$, where $\rho$ is the density of the working fluid, $\tau_{p}$ is the particle viscous relaxation time and $\tau_{\eta}$ is the Kolmogorov timescale, so that the particles reliably follow the fluid motion \citep{xu2008motion}. However, given the initial and time-varying density field, the primary challenge for stratified flows is reducing the settling velocity of the particles 
\begin{equation}
V_p = \frac{g(\rho_p - \rho)d_p^{2}}{18\rho\nu},
\end{equation}
so that they remain suspended prior to beginning the experiment and also during the experiment itself. 

Having seeded the flow, a thin slice of thickness $\delta z$ ($O(1)$ mm) is illuminated by a light sheet centred at $z_c$ in world coordinates and imaged using a camera. The three-dimensional (3D) illuminated slice is projected onto the plane of the two-dimensional (2D) camera sensor
\begin{equation}
\mathbf{X} = \mathbf{P}(\mathbf{x}),\label{2DProject}
\end{equation}
where
\begin{equation}
\mathbf{X}  = \begin{pmatrix}
X \\
Y \end{pmatrix}   
\quad\textrm{and}\quad
\mathbf{x} = \begin{pmatrix}
x \\
y \\ 
z = z_c \\ \end{pmatrix}.
\end{equation}
Here $\mathbf{X}=(X,Y)$ are the sensor coordinates of the camera, $\mathbf{P}$ is the projection function that maps from the 3D image onto the 2D camera sensor, and $\mathbf{x}=(x,y,z=z_c)$ are the world coordinates at the centre of the slice being illuminated. Similarly, an inverse mapping $\mathbf{P}^{-1}$ can be determined to map between pixel coordinates $\mathbf{X}$ and world coordinates $\mathbf{x}$
\begin{equation}
\mathbf{x} = \mathbf{P}^{-1}(\mathbf{X}),
\end{equation}
for a given $z_c$. To determine the velocity field, a sequence of images containing particles are captured. For PTV, individual particles are tracked over the sequence of images (for example, \cite{adrian1991particle}, \cite{dalziel1993rayleigh}, and \cite{schanz2016shake}). For PIV, the focus herein, the raw particle images are subdivided into small interrogation windows and compared over a time period $\Delta t$ (either the time between light sheets for a pulsed source or the time between exposures for a continuous source). At the core of any PIV algorithm, this initial comparison is often a pixel-accurate pattern matching function (e.g. a 2D cross-correlation) that determines the optimal image displacement $\Delta \mathbf{X}$ between two interrogation windows separated by a time $\Delta t$. It is worth noting that most PIV algorithms have additional steps that allow for subpixel accuracy but still inevitably rely on the initial pattern matching procedure \citep{adrian2005twenty}.

Using the optimal displacement $\Delta \mathbf{X}$, the velocity field in pixel coordinates $\mathbf{U}=\frac{\Delta \mathbf{X}}{\Delta t}$ can be mapped, using the projection function $\mathbf{P}^{-1}$, to world coordinates
\begin{equation}
{\mathbf{U}\left({\mathbf{X},z_c}\right)} \xmapsto{\quad \mathbf{P}^{-1}(\mathbf{X}) \quad} \hat{\mathbf{U}}
(\mathbf{x}).
\end{equation}
{Note that the units of $\hat{\mathbf{U}}
(\mathbf{x})$ are still in pixels per unit time.} To compute the velocity field in world units {per unit time}, knowledge of the mapping derivatives is also required through the Jacobian $\mathbf{J}^{-1}=\frac{\partial\mathbf{P}^{-1}}{\partial\mathbf{X}}$ (here a $2 \times 2$ matrix) with {with each element expressed as world units per pixel}. This is also mapped to world coordinates 
\begin{equation}
\mathbf{J}^{-1}(\mathbf{X}, z_c) \xmapsto{\quad \mathbf{P}^{-1}(\mathbf{X}) \quad} \hat{\mathbf{J}}^{-1}(\mathbf{x}),
\end{equation}
{noting} that $\hat{\mathbf{J}}$ still has units of world/pixel. It follows that the second-order accurate velocity field in world units (and in world coordinates) at time $t$ can then be calculated from
\begin{equation}
\hat{\mathbf{u}}_{2D} \simeq \hat{\mathbf{J}}^{-1}\hat{\mathbf{U}},\label{2D2CCalc}
\end{equation}
where $\hat{\mathbf{u}}_{2D} = (u, v)^\intercal$ as only the projected in-plane measurements can be measured in traditional planar PIV.

Given the pattern matching required, PIV generally relies on a light sheet that is thick relative to the particles displacement in the $z$ direction (the direction normal to the light sheet) in the time $\Delta t$ so the majority of particles remain within the illuminated volume between images (we shall return to this in \S\ref{meanMotion}).

Although still useful, these 2D measurements only yield two-component velocity data of the thin, approximately 2D illuminated slice (2C2D) and there is an increasing demand to capture all components of velocity, especially important in turbulent flows that are inherently 3D. The stereo PIV method  \citep{prasad2000stereoscopic} allows all three components of velocity ($u$,$v$,$w$) to be calculated on the illuminated plane by introducing a second camera (Camera B) viewing the flow with a different perspective from the first camera (Camera A). The two perspectives of the illuminated slice allow the $z$ component (the component in the direction normal to the plane of the light sheet) of velocity $w$ to be calculated giving all three components of velocity on the plane (2D3C). 

\subsection{Volumetric Velocity Measurements}

Planar PIV techniques (2D2C and 2D3C) only allow measurements within a single slice. However, over the past decade or so there has been a lot of development in volumetric velocity field measurement techniques. A full review of techniques is beyond the scope of this paper and the reader is referred to \citet{scarano2012tomographic} for further discussion. However, to justify the methodology described in this paper, we briefly summarise the two approaches commonly used to capture a velocity field in a volume: tomographic PIV and scanning PIV (PIV-S). 

Tomographic PIV makes use of an expanded illuminated region, typically produced by a laser, such that a volume is illuminated. The flow is seeded with passive tracer particles, as in standard PIV, and imaged by multiple cameras with different perspectives. By comparing two instants in time (i.e. two illuminated volumes) the multiple views from the cameras are combined, through careful calibration, and a three-component velocity field in the volume can be inferred \citep{scarano2012tomographic}. 

A challenge when making optical measurements such as PTV or PIV in stratified flows is dealing with refractive index variations. Whereas some measurement techniques in stratified flows make use of the refractive index variations within the flow (e.g. synthetic schlieren \citep{dalziel1998} and shadowgraph techniques \citep{hesselink1988digital}), such variations present a real obstacle for PIV measurements. PIV relies on being able to image distinct particle patterns and accurately determine their location and displacement between pairs of images. This becomes difficult when strong refractive index variations are present within the flow, especially when, due to fluid motion, the refractive index of the fluid between the particle pattern and the camera can vary with time. As well as producing an error in the absolute position of the particle pattern (and hence the velocity vector), if the refractive index is varying on a timescale comparable to the $\Delta t$ between frames, the true displacement of the particle pattern (the magnitude and direction of the velocity vector) is compromised \citep{dalziel2007simultaneous}. Moreover, if strong refractive index variations are present between the camera and the light sheet it can be a challenge to even image such small particles making it impractical to run a PIV algorithm. Fortunately, these unwanted effects can be minimised by choosing two solutions that have different densities but the same refractive index (e.g. \citet{mcdougall1979elimination} and \citet{dalziel1999self}). 

Unfortunately, the tomographic approach is more susceptible to any residual refractive index mismatch than the planar measurements making it less suited to the study of stratified flows. This is due to the fact that the tomographic approach relies on accurately positioning particles, triangulating them between the multiple cameras imaging them, and then calculating the individual particles (or volume of particles) 3D displacement at subsequent times. Moreover, residual refractive index variations lead to lower particle yields for most algorithms, particularly for particles with 3D locations more remote from the cameras. In comparison, the planar PIV approach is more robust to such errors as the pattern matching algorithm is less sensitive to absolute refractive index variations than to their rate of change as shown in \citet{dalziel2007simultaneous}. 

As well as the need for multiple cameras (typically four or more), spatial resolution is also a limiting factor for a tomographic system as all of the particles throughout the measurement volume have to be distinguishable simultaneously in the camera images. This means the yield of vectors in the total volume is limited, with either the spatial resolution or spatial extent being compromised. The resolution of a tomographic system is typically expressed in particles per pixel (ppp). The latest algorithms can achieve 0.1 ppp yielding one velocity vector per particle \citep{schanz2016shake}. To compare this with {planar PIV, consider} a resolution of one velocity vector every 8 x 8 pixels, a 4 megapixel camera will give $\sim 6\textrm{\:x\:}10^4$ vectors on the plane for PIV compared with $\sim 4\textrm{\:x\:}10^5$ vectors for the 0.1 ppp tomographic system. However, by building a volume of more than six planes the PIV approach produces a higher volumetric vector yield. On the other hand, an advantage to the tomographic approach is that the 3D field is {effectively} instantaneous, i.e. there is no time lag when imaging across the volume and the flow is frozen by the short $O(10)$ ns pulse of the laser. 

An alternative method to achieving volumetric measurements is PIV-S. In PIV-S, a light sheet is either rapidly scanned through a volume seeded with particles, producing a volume of particles as seen by one camera (volumetric approach), or planar (2C2D or 3C2D) measurements are made at discrete locations throughout the volume (planar approach). 

In the volumetric approach, the reconstructed volume can be broken down into subvolumes of particles (voxels) and pattern matched in 3D to yield the full three-component (3C3D) velocity field. This scanning approach has advantages over tomographic PIV as only one camera is needed, but is limited in application due to the rapid scanning speed needed for the light sheet, to effectively freeze the flow in the volume instantaneously. For typical flows of interest, this method requires a high-speed pulsed laser (or powerful continuous wave laser) and a high-speed camera ($O(10^4)$ fps) that currently limits the spatial resolution of the measurements. 

The planar approach dates back to \citet{brucker1995digital} who used a large mirror on a stepper motor to position a light sheet at discrete locations throughout the measurement volume. Essentially, a series of images were captured at the various discrete locations, which enabled 2D PIV to be calculated on the various subvolumes that were stacked, after processing, to reveal the 3D structure of the flow. This approach is limited in quantifying faster flows, as the scanning over the volume was relatively slow, due to the rotation of a mirror large enough to accommodate the light sheet. More recent studies have opted to scan a light sheet using oscillating mirrors positioned using galvanometers (for accurate and fast positioning). {For example, \citet{krug2014combined} produced a light sheet by passing a laser beam through a cylindrical lens before scanning the sheet using an oscillating mirror. The light sheet, while scanning, passed through a second cylindrical lens to further expand the sheet. Ultimately, this approach produced slowly diverging rather than parallel light sheets (as a small angular adjustment was needed to sweep the whole volume of interest) across the volume.} Unfortunately, this approach is typically limited in spatial extent as it requires large, high-quality optics so volumes are typically only a few centimeters deep.

A novel alternative approach was demonstrated by \citet{olsthoorn2017three} where the light sheet remained at a fixed position but the experiment was effectively moved (in this case by displacing the density interface relative to the light sheet). Multiple identical experiments were conducted and stereo PIV measurements taken, with the relative light sheet position moving between each experiment. In post processing, the 2D3C velocity fields of each experiment were stacked, with each experiment corresponding to a different position of the light sheet in the frame of reference of the density interface, to produce an ensemble 3D velocity field. However, this method is quite laborious, as many identical experiments need to be performed to capture one 3D field, and ultimately only the reproducible component of the velocity can be obtained. 

\begin{figure}[h!]
\centering
\includegraphics[width=0.95\textwidth]{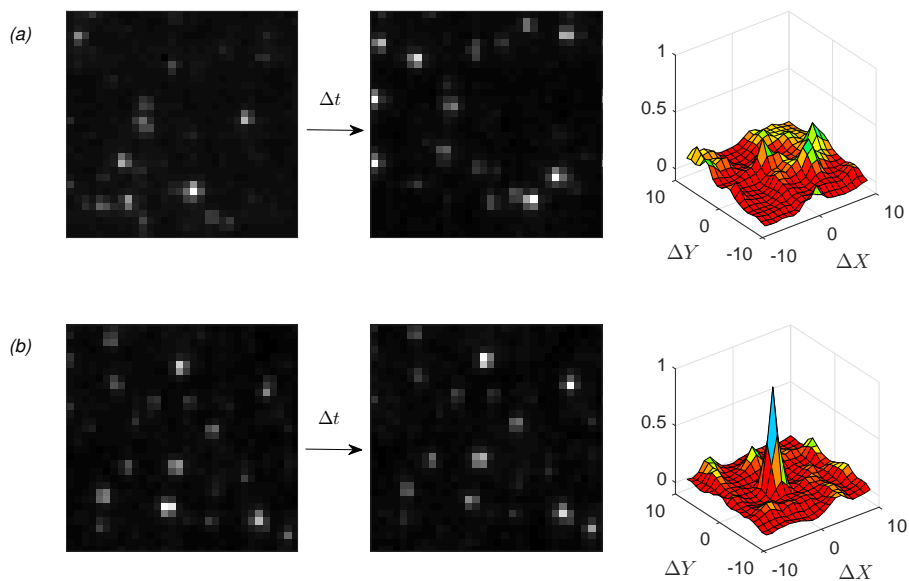}
\caption{Particle pattern in two interrogation windows separated by a time $\Delta t$ for (a) non-overlapping light sheets and (b) overlapping light sheets. For both, the normalised cross-correlation signal is shown on the right. In (a) a recognisable particle pattern is not visible between the two interrogation windows as evidenced by the noisy cross-correlation signal. In (b) the particle pattern is detectable by eye and is evident by the clear peak in the normalised cross-correlation indicating a pixel shift of $\Delta X = 0$ and $\Delta Y = 1$}\label{corrMirrorsNoMirrors}
\end{figure}

The fundamental issue when using the scanning light sheet approach is ensuring sufficient light sheet overlap between a pair of images to permit accurate velocity measurements. For PIV, if the majority of particles within an interrogation window are lost in the time $\Delta t$, either due to high out-of-plane velocities compared with $\delta z/\Delta t$ or, in the scanning system, because the light sheet has moved, the matching function will not yield useful information about the flow. This is illustrated in figure \ref{corrMirrorsNoMirrors}, which shows an interrogation window separated by $\Delta t$ from the same location in two nominally identical experiments {with the same (mean) scanning speed}. In figure \ref{corrMirrorsNoMirrors}a, a traditional continuous scanning approach was used and there is a poor overlap of illuminated slices. Therefore, {the relationship between the patterns produced by the particles is not clear} between snapshots. {In contrast}, in figure \ref{corrMirrorsNoMirrors}b our new scanning approach was used and a good overlap of illuminated slices was achieved. Here, a clear particle pattern is recognised between snapshots. Inevitably, where there is poor overlap the resultant matching function (a 2D cross-correlation in figure \ref{corrMirrorsNoMirrors}) is noisy. In contrast, where good overlap is achieved there is a clear peak in the matching function and PIV analysis can be reliably performed.

In a traditional scanning system, considering a light sheet of thickness $\delta z$ traversing at a constant velocity $V$, it is clear that there will be no light sheet overlap when
\begin{equation}
	\delta z < V\Delta t.\label{constraint}
\end{equation}
This puts a large constraint on the scanning approach as either the light sheet has to be excessively thick (reducing the resolution in that direction) or the scanning speed slow (reducing the temporal resolution of the measurements). Unfortunately, it is not practical to change $V$ rapidly by starting and stopping a traverse due to the inertia of the traverse system when it must also move $\sim 2$ mm in $\sim 10$ ms (the typical interval between pairs of laser pulses) and traverse a total distance $O(10)$ cm for a typical experiment. Simply starting and stopping the traverse in this manner would require a powerful motor and introduce undesirable vibrations into the traverse carriage that would compromise the positioning of the light sheet during the scanning sequence. The new scanning method detailed in this paper allows the constraint \eqref{constraint} to be relaxed, enabling both thin light sheets and fast scan speeds $V$.   

\subsection{Density Measurements}

Density fields are often measured using planar laser induced fluorescence (PLIF). In a standard PLIF set-up, a fluorescent dye is added to one of the solutions that you wish to passively tag and, through careful calibration, the concentration field of the fluoresced dye can be related to the underlying density field \citep{crimaldi2008planar}. By illuminating a thin slice and projecting the illuminated volume onto the plane of the camera sensor, 2D planar measurements of the density field can be obtained. As with PIV measurements, care has to be taken to match the refractive indices of the solutions used so that density measurements can be accurately positioned and sharp gradients maintained (i.e. no loss of focus or blurring due to refractive index variations along the light path and no focusing/defocusing of the light sheet).

Simultaneous velocity and density measurements can be achieved through a careful choice of fluorescent dye, selecting the absorption wavelength of the dye to the wavelength of the laser, and using a dye with a large Stokes shift, such that the fluoresced wavelength is significantly larger than the absorption wavelength (e.g. \citet{webster2001simultaneous}). This allows the scattered light from the particles to be filtered out enabling different cameras to simultaneously obtain velocity measurements using PIV and density measurements using PLIF. Typically, Rhodamine 6G is used as the fluorescent dye and illuminated with, for example, a frequency doubled Nd:YAG or Nd:YLF laser. 

As with the velocity measurements, density measurements can be made over a volume using an expanded light sheet and a tomographic approach. However, tomographic approaches to determine complex density fields currently rely on a prohibitively large number of distinct views of the illuminated volume and do not represent a viable approach for turbulent flows. The approach can be utilised where it is feasible to use ensemble data (e.g. \citet{hazewinkel2011tomographic}), but this is typically not the case for many flows of interest. Fortunately, straightforward scanning approaches offer a viable alternative.

As PLIF only requires a single image for each plane, it is relatively straightforward to extend it to 3D by scanning a planar light sheet over a volume. Numerous studies have made such measurements (e.g. \citet{dahm1991direct} and \citet{tian20033d}) but, to date, simultaneous velocity and density measurements in a volume have typically been limited to small spatial extents (e.g. \citet{krug2014combined}). As it is desirable to have simultaneous density and velocity measurements over large volumes, we aim to extend the PIV-S approach to allow fast scanning of large volumes that readily allow the simultaneous capture of 3D density fields.  

\section{Methodology}\label{method}

\subsection{Hardware}

\begin{figure}[h!]
\centering
\includegraphics[width =  0.75\textwidth]{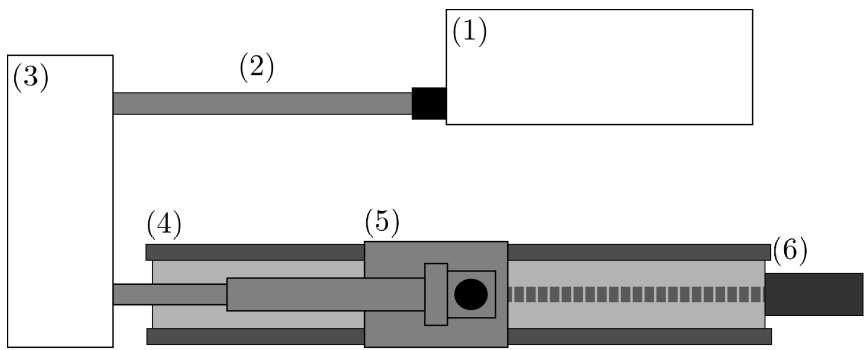}
\caption{Schematic of the scanning PIV/PLIF system showing a plan view of the scanning apparatus: (1) laser source; (2) beam tube to enclose the laser beam; (3) housing containing the oscillating mirror assembly that allows the beam to be displaced vertically; (4) linear guide rail that allows the light sheet-producing-optics to travel in the scanning direction; (5) carriage containing the light-sheet-producing optics; and (6) motor used to move the traverse}\label{LaserSchemPlan}
\end{figure}

In the present study, a pulsed, dual-cavity laser is used as the light source with a maximum repetition rate of $f_{laser}$ for each cavity. To accomplish the scanning, a mirror and the light-sheet-producing optics, consisting of a set of diverging (plano-concave) cylindrical lenses, are positioned on a linear traverse. The layout of all of the components that produce and scan the light sheet is shown in figure \ref{LaserSchemPlan}. 

\begin{figure}[h!]
\centering
\begin{subfigure}[t]{0.3\textwidth}
\includegraphics[width=\textwidth]{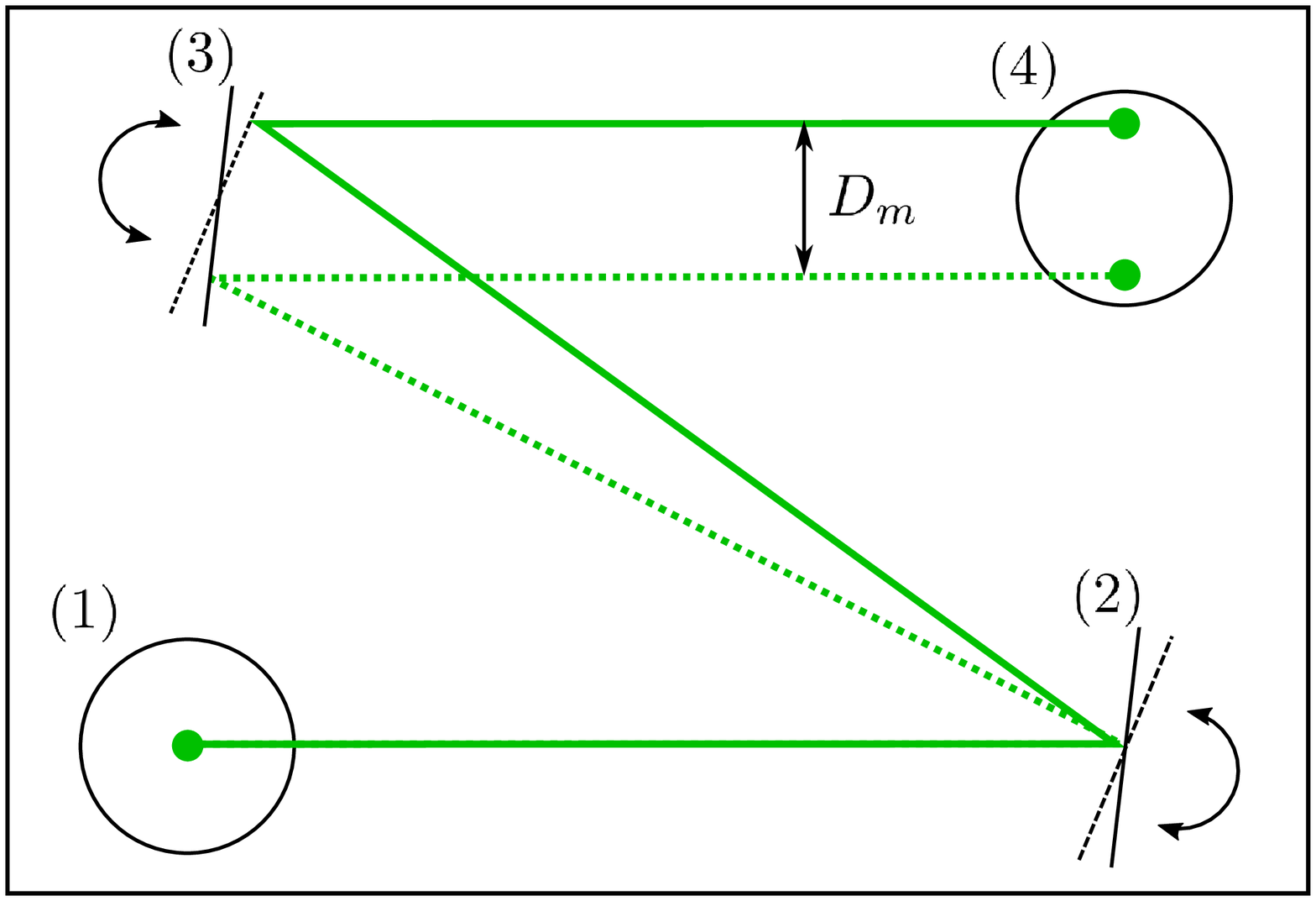}
\caption{Front view of mirror assembly} \label{fig:MirrorsSchem_a}
\end{subfigure}       
\begin{subfigure}[t]{0.45\textwidth}
\hspace{0.25cm}
\includegraphics[width=\textwidth]{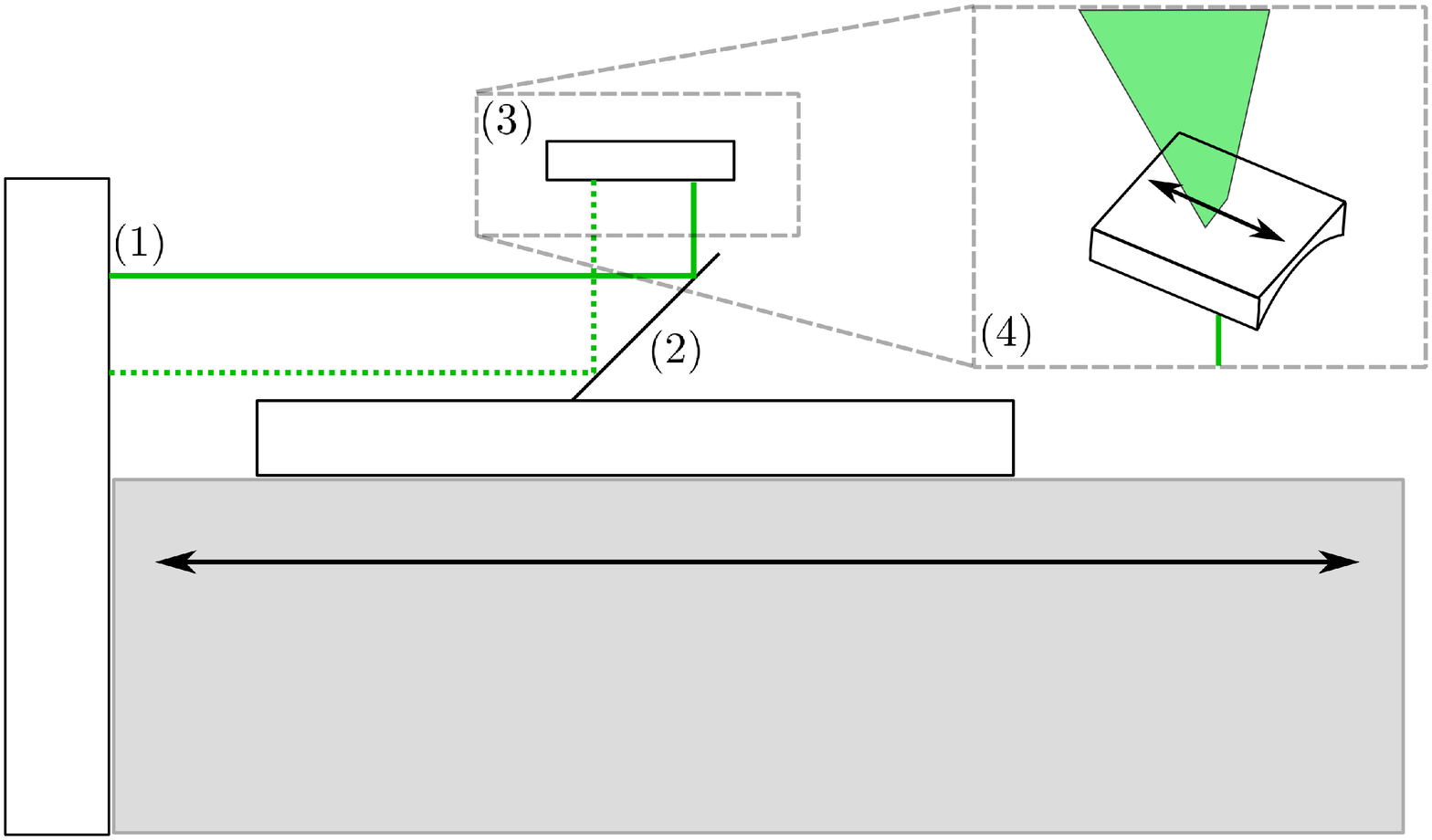}
\caption{Side view of beam path after exiting the mirror assembly} \label{fig:MirrorsSchem_b}
\end{subfigure}    
\caption{{(a) Details of the oscillating mirror assembly: (1) first fixed 45$^{\circ}$ mirror that deflects the incoming beam from the laser into the plane of the sketch; (2) and (3) oscillating mirrors that allow the beam to be displaced vertically a distance $D_m$; (4) second fixed 45$^{\circ}$  mirror that sends the beam down to the scanning optics. (b) Beam path after exiting the oscillating mirror assembly, looking from the side of (a): (1) beam after exiting the mirror assembly in (a); (2) fixed 45$^\circ$ mirror on the traverse carriage that makes the laser beam vertical (and the vertical beam displacement horizontal); (3) cylindrical lenses to produces a light sheet; (4) 3D views of (3) to illustrate the curvature of the cylindrical lens}}\label{LaserMirrors}
\end{figure}

The novel addition to the set-up is a pair of oscillating mirrors that can be used to superimpose a small additional translation of the light sheet in the scanning direction. To accomplish this, the beam from the pulsed laser is first reflected by a fixed 45$^\circ$ mirror onto the first oscillating mirror. The beam is reflected by this mirror onto the second oscillating mirror before being redirected to the traverse carriage by a second fixed 45$^\circ$ (see figure \ref{fig:MirrorsSchem_a}). Once on the traverse carriage, the beam is reflected by a final 45$^\circ$ mirror that passes the unexpanded beam through a set of diverging cylindrical lenses to form a light sheet perpendicular to the traversing direction (see figure \ref{fig:MirrorsSchem_b}). The new scanning method makes use of the fact that, by construction, cylindrical lenses have no curvature in the direction normal to the plane of the light sheet they produce. Therefore, by translating the incoming beam in this direction, the oscillating mirrors can offset the light sheet in the direction of scanning independently of the position of the traverse. Furthermore, if the translation keeps the beam parallel, then the light sheets will be parallel. By positioning the oscillating mirrors as shown in figure \ref{fig:MirrorsSchem_a}, and rotating them in tandem, the beam is displaced vertically before entering the light-sheet-producing optics. The parallel displacement in the vertical, after passing through the 45$^{\circ}$  mirror at the base of the optics (see figure \ref{fig:MirrorsSchem_b}), results in a parallel displacement in the traversing direction. To preserve the parallel nature of the beam displacement, the thickness of the light sheet $\delta z$ should be controlled using a focusing module (e.g. a set of spherical lenses) between the laser and the oscillating mirror assembly.  

\begin{figure}[h!]
\centering
\includegraphics[width = 0.4\textwidth]{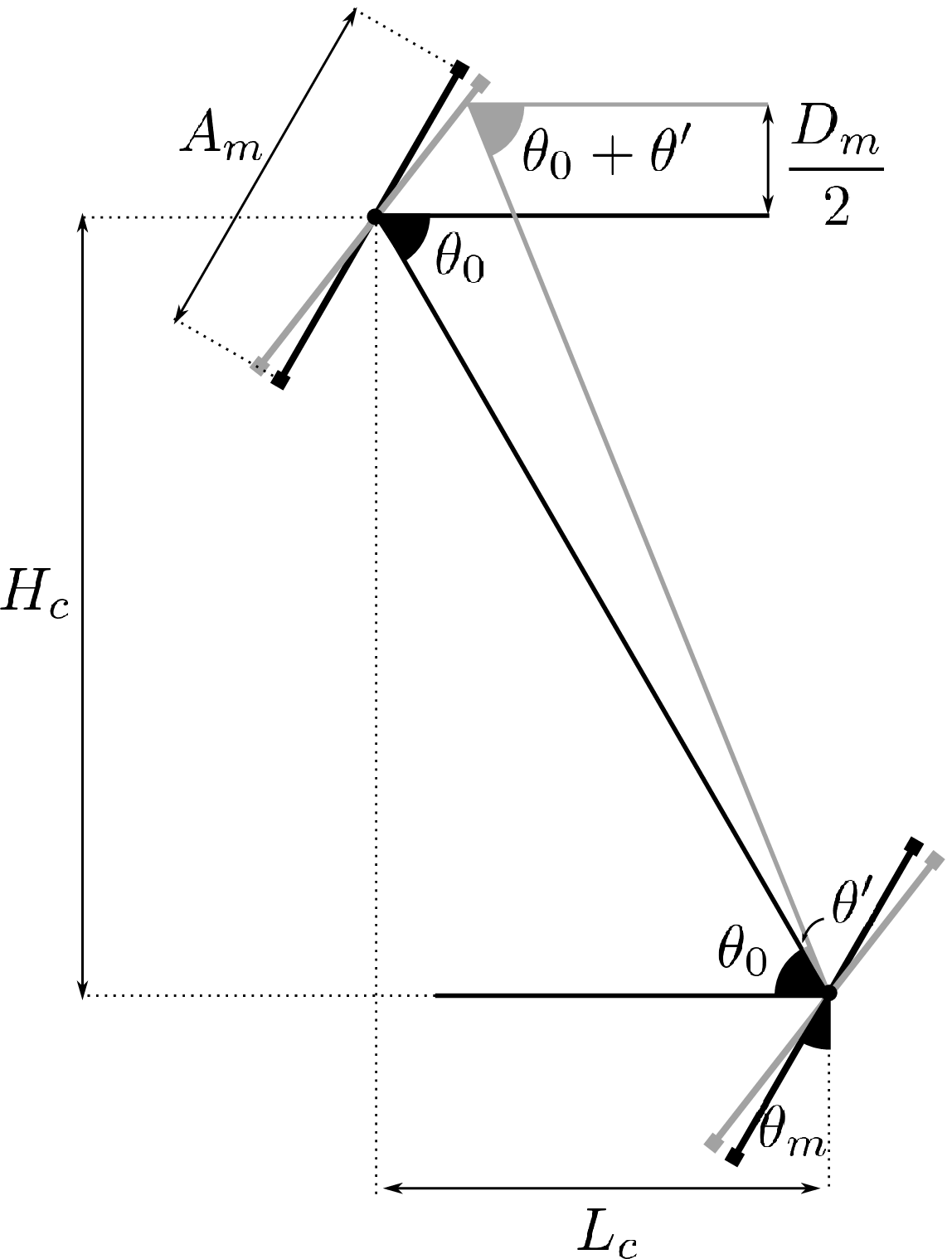}
\caption{Schematic showing the layout of the oscillating mirrors illustrating the positioning and subsequent rotation to achieve a vertically displaced beam} \label{MirrorCalcFig}
\end{figure}

\begin{figure}[h!]
    \centering
    \begin{subfigure}[b]{0.45\textwidth}
        \includegraphics[width=\textwidth]{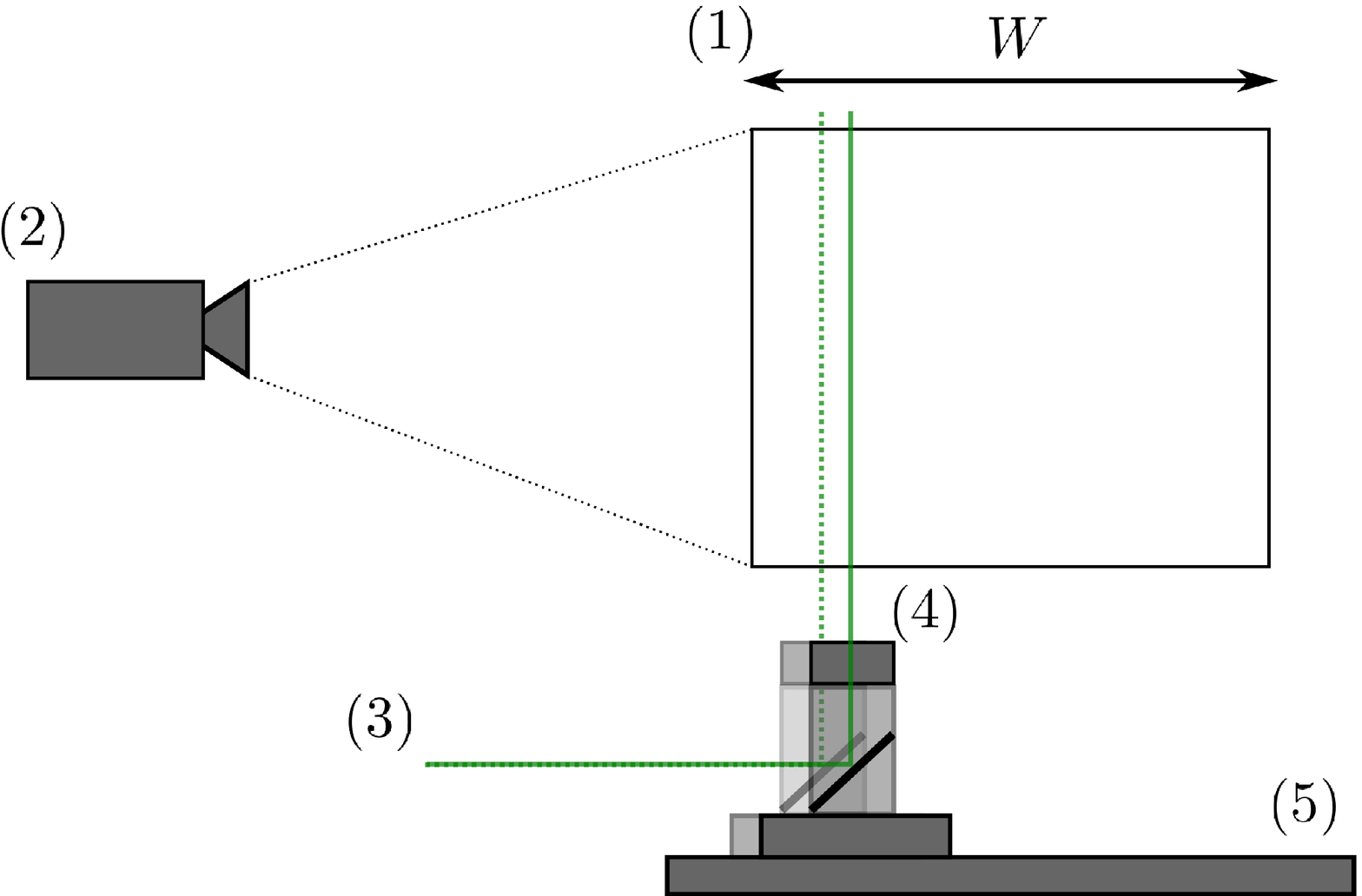}
        \caption{Without oscillating mirrors}
        \label{LaserSchemSideNoMirror}
    \end{subfigure}
     \begin{subfigure}[b]{0.45\textwidth}
        \includegraphics[width=\textwidth]{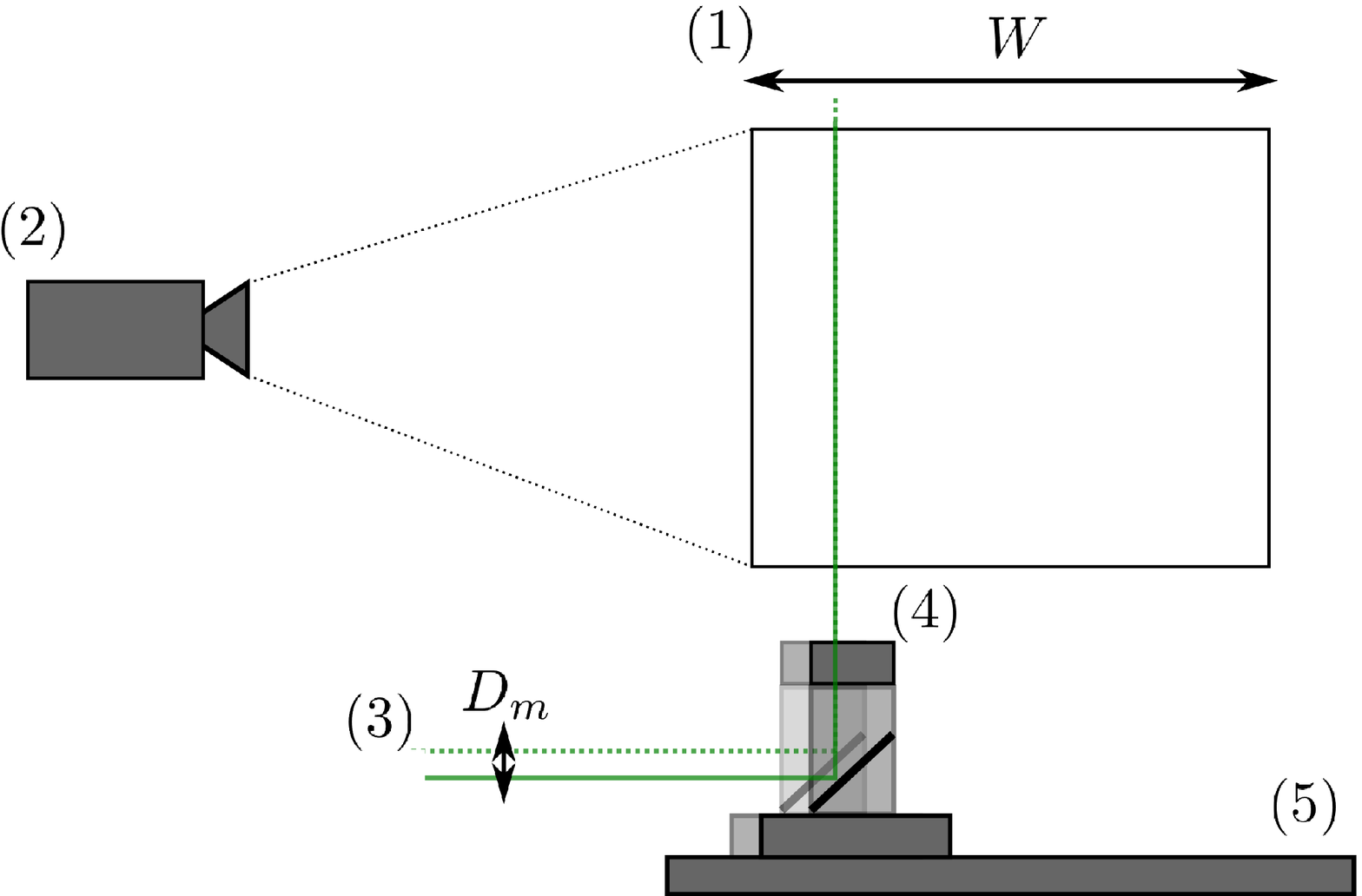}
         \caption{With oscillating mirrors}
        \label{LaserSchemSide}
    \end{subfigure}
    \caption{Side view of the scanning PIV/LIF system (a) without and (b) with oscillating mirrors: (1) region of interest; (2) bank of three cameras; (3) incoming beams at time $t_0$ (dotted) and $t_0 + \Delta t$ (solid); (4) 45$^{\circ}$ mirror mounted on the traverse that then passes the beam through light-sheet-producing optics (all faint at time $t_0$). The vertical shift $D_m$ of the beam in time $\Delta t$ on the left schematic allows two beams to be coincident within the region of interest for PIV measurements. Finally, (5) shows the linear guide rail that allows the optics to be scanned a distance $W$}\label{LaserSchemSideAll}
\end{figure}

For a given oscillating mirror assembly (i.e. a prescribed resting state angle $\theta_0$ along with both $L_c$ and $H_c$ the horizontal and vertical distance between the mirror centres, respectively) the vertical beam displacement $D_m/2$ can be determined from
\begin{equation}
\frac{D_m}{2}=L_c\sin(\theta_0 + \theta') - H_c\cos(\theta_0 + \theta').
\end{equation}
Here, $\theta'$ is the angular perturbation away from the resting state $\theta_0$ (see figure \ref{MirrorCalcFig}). The apertures $A_m$ of the two oscillating mirrors do not need to be equal and, as clear from figures \ref{fig:MirrorsSchem_a} and \ref{MirrorCalcFig}, the aperture of the first mirror in the sequence only needs to be large enough to accommodate the incoming beam. The aperture of the second mirror needs to be larger than the beam diameter to accommodate the beam displacement, and its size (along with the resting angle $\theta_0$) will determine the maximum beam displacement possible. It is useful to note that the angular perturbation $\theta'$ is twice the angular perturbation given to the mirror itself (i.e. twice the mechanical angle as $\theta_0 = 2\theta_m$). Without the oscillating mirrors, it would not be possible to get precisely overlapping light sheets while scanning the light-sheet-producing optics, as illustrated in figure \ref{LaserSchemSideNoMirror}. Figure \ref{LaserSchemSideNoMirror} shows a side view of the scanning system without the oscillating mirror assembly, demonstrating that sequential light sheets do not overlap. Where as, in figure \ref{LaserSchemSide} the oscillating mirror assembly is used displace the laser beam vertically between the two laser pulses producing two spatially coincident light sheets.

\begin{figure}[h!]
\centering
\includegraphics[width = 0.6\textwidth]{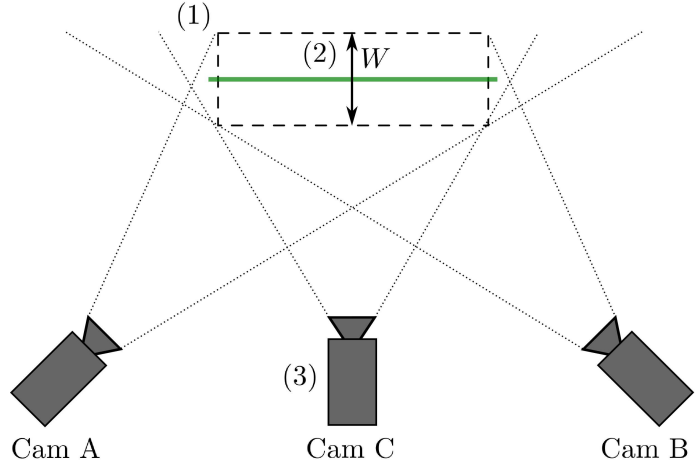}
\caption{Schematic of the scanning PIV/PLIF system showing a top view of the system and the camera layout: (1) the region of interest; (2) the light sheet that is scanned a distance $W$; (3) the array of cameras: Cam A and B for particle images and Cam C for the concentration of a scalar field for PLIF.}\label{LaserSchemCam}
\end{figure}

To simultaneously obtain the full velocity field and density field, a set of three cameras are required for the measurements with a typical layout shown in figure \ref{LaserSchemCam}. For the present discussion, we keep the cameras stationary as this removes a potential source of noise from the velocity fields (see \S\ref{multTraverses}). As is the case for tomographic PIV, having the cameras stationary requires that the lenses have a sufficiently small aperture (large f-number) so that the particles remain adequately focused across the volume of interest. For the stereo PIV measurements, two cameras are typically positioned with some angular separation and fitted with Scheimpflug adapters to provide better focusing on the scanned planes. These cameras need to be fitted with shortpass filters (or, preferably, bandpass filters centred at the wavelength of the laser) to eliminate the signal from the fluoresced dye. For the density measurements, the third camera is ideally positioned with its optical axis normal to the light sheet (to minimise distortion) and needs to be fitted with a longpass filter to remove the light directly scattered from the particles, leaving only the signal from the fluorescent dye.

\begin{figure*}[h!]
\centering
\includegraphics[width =  0.9\textwidth]{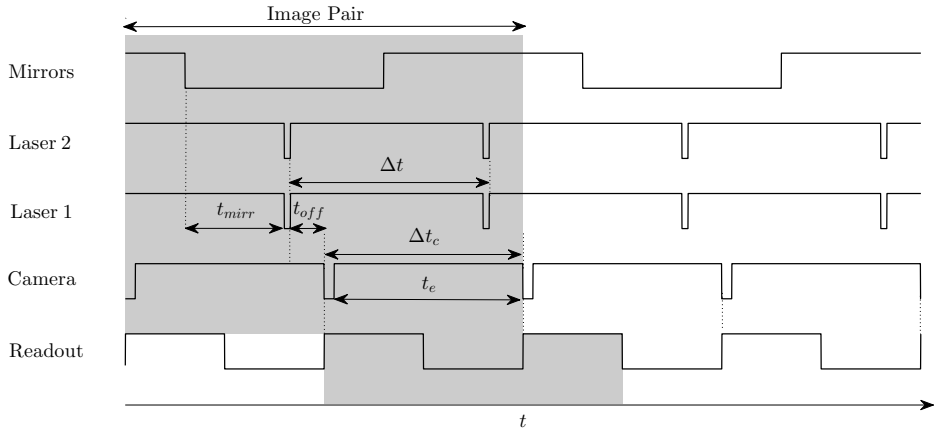}
\caption{Schematic of the timing sequence sent to the various components of the system when in single-pulse mode. The laser pulse is triggered from the rising edge (+ve) of the pulse train to the lasers. An image pair (one plane of velocity) is indicated by the shaded region}\label{pulseTrain_SP}
\end{figure*}

\begin{figure*}[h!]
\centering
\includegraphics[width =  0.9\textwidth]{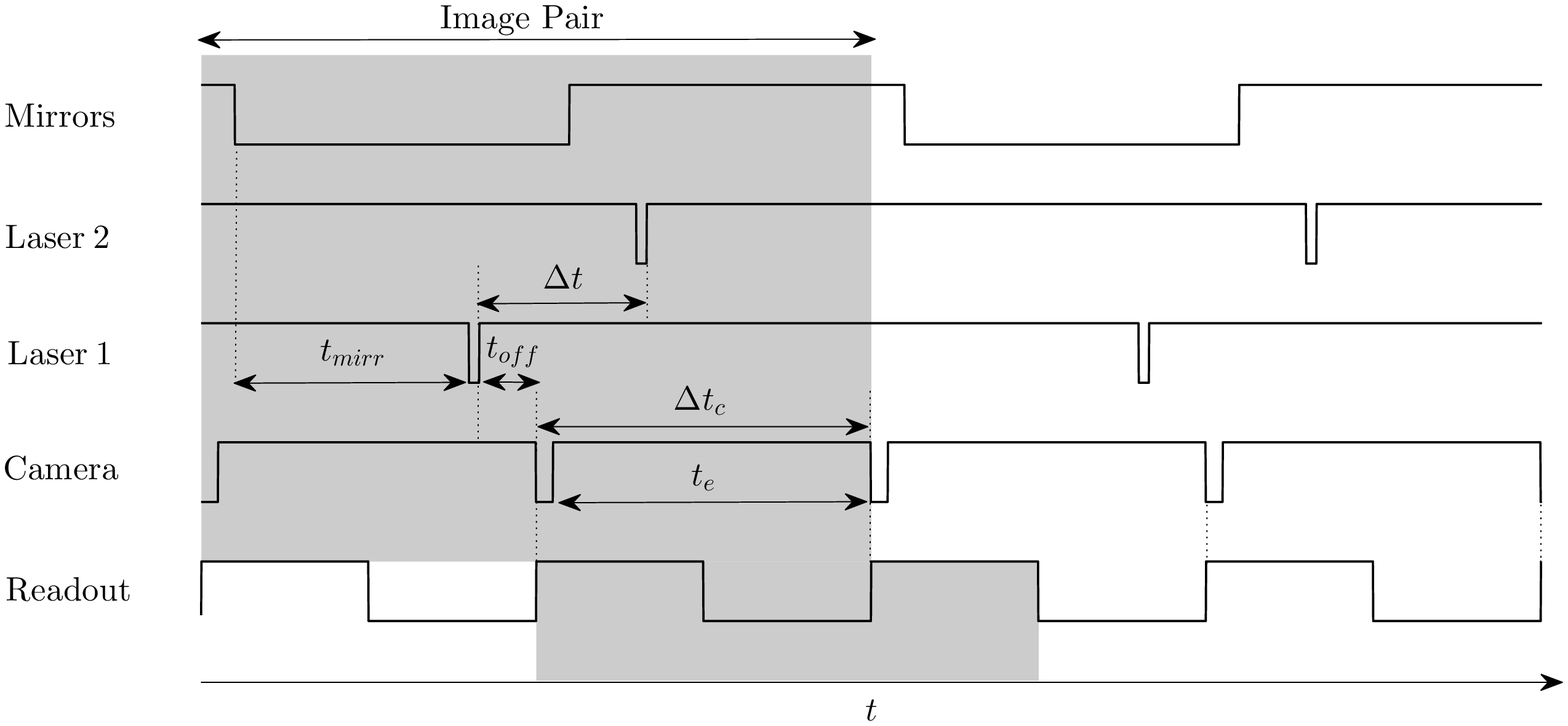}
\caption{Schematic of the timing sequence sent to the various components of the system when in double-pulse mode. The laser pulse is triggered from the rising edge (+ve) of the pulse train to the lasers. An image pair (one plane of velocity) is indicated by the shaded region}\label{pulseTrain_DP}
\end{figure*}

The pulsed laser, oscillating mirrors, traverse and cameras are all triggered using a hardware-based timing system with all of the components sharing a common clock to keep everything synchronised. Typical pulse sequences for the various components are shown in figures \ref{pulseTrain_SP} and \ref{pulseTrain_DP}. Figure \ref{pulseTrain_SP} shows the pulse sequence in single-pulse mode (firing the two laser cavities together) and figure \ref{pulseTrain_DP} shows the double-pulse mode (laser cavities are fired out of phase). As is generally the case for PIV, firing the lasers in double-pulse mode allows a smaller effective $\Delta t$ between light sheets and enables fast flows to be measured even if they would exceed an optimal particle displacement at the cameras' maximum frame rate or the pulsed lasers maximum repetition rate. Conversely, the double-pulse mode can be used to image slow flows at a relatively high frame rate by positioning the laser pulses at the beginning and end of sequential frames. The signals to the cameras control the exposure time and image readout (triggered by the negative edge) so the cameras' global shutters are open for a time $t_e$. Given that the flow field is not illuminated except for the short (typically $O(10)$ ns) pulse from the lasers, the effective timing between images captured is determined purely by the laser pulse timing. In single-pulse mode, the maximum temporal resolution of the measurements is determined by the smaller of $f_{laser}$ and the maximum frame rate of the cameras. In double-pulse mode, the maximum temporal resolution is determined by the smaller of $2f_{laser}$ and the maximum frame rate of the cameras. In the scanning system (in either pulse mode), the time between each measurement subvolume is $2\Delta t_c$, with $\Delta t_c$ the time between camera frames.

The signal to the mirrors is a square wave alternating between the two mirror positions to translate the light sheet (forward and back) to compensate for the traverse motion. As, in practice, the inertia of the oscillating mirrors means their orientations cannot change instantaneously, the phase of the mirror signal is set so that the mirrors can settle in their required orientations prior to the laser being pulsed. In other words, the position of the laser pulse has to fall somewhere between the start of the current mirror position and the end of the current frame, i.e. $t_{mirr}\neq0$ and $t_{off}\neq0$ as shown in figures \ref{pulseTrain_SP} and \ref{pulseTrain_DP}. In practice, the smallest allowable $t_{mirr}$ will depend on the oscillating mirrors used as well as the distance between the two mirrors to ensure the mirrors have enough time to reach their steady position. A smoother waveform, such as a sinusoid, timed so that the laser pulse occurred at the correct moment of the waveform would allow smaller usable $t_{mirr}$ but a square wave is often sufficient and simpler to implement. Furthermore, by increasing the separation between the mirrors, a smaller rotation of the mirrors is needed to achieve the same beam displacement, and thus the mirrors require less time to stabilise their orientation. However, a compromise is needed as reducing the rotation required increases the error of the beam displacement due to the inherent accuracy of the oscillating mirror rotation.

\begin{figure*}[h!]
\centering
\includegraphics[width =  0.9\textwidth]{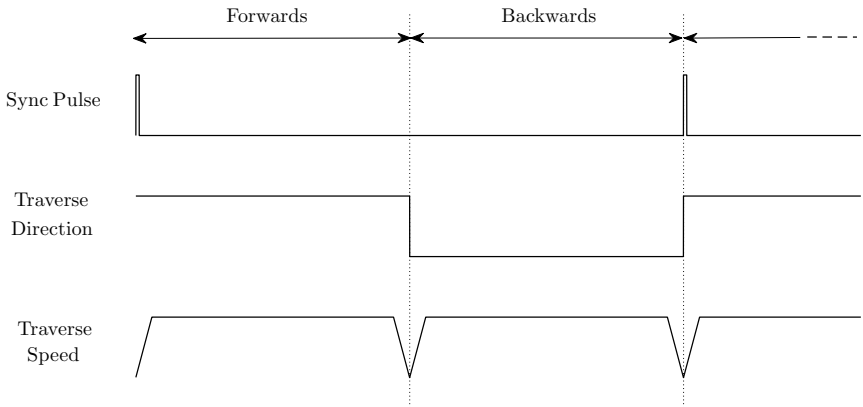}
\caption{Schematic showing the timing signal over the course of a complete scan. Acceleration/deceleration phases are shown for the traverse, along with the direction signal that controls the direction of travel for the traverse. A sync pulse is shown that is used to trigger the recording of a sequence of images for all cameras at the start a scan}\label{pulseTrain_bigPic}
\end{figure*}

Typical signals of the traverse speed, traverse direction and sync pulse over a complete scan are shown in figure \ref{pulseTrain_bigPic}. The signals are periodic and allow the traverse carriage holding the optics that produce the light sheet to continuously sweep back and forth through the measurement volume. The `direction' signal controls the direction of the traverse, and the `scan sync' signal is pulsed at the start of every forward scan to initiate the capture of a series of images on all cameras. Note that the scan sync pulse only initiates saving on all cameras, to ensure all recorded sequences start at the beginning of the forward scan, but is periodic allowing, for example, every $n$th scan to be saved.

\begin{figure*}[h!]
\centering
\includegraphics[width =  0.9\textwidth]{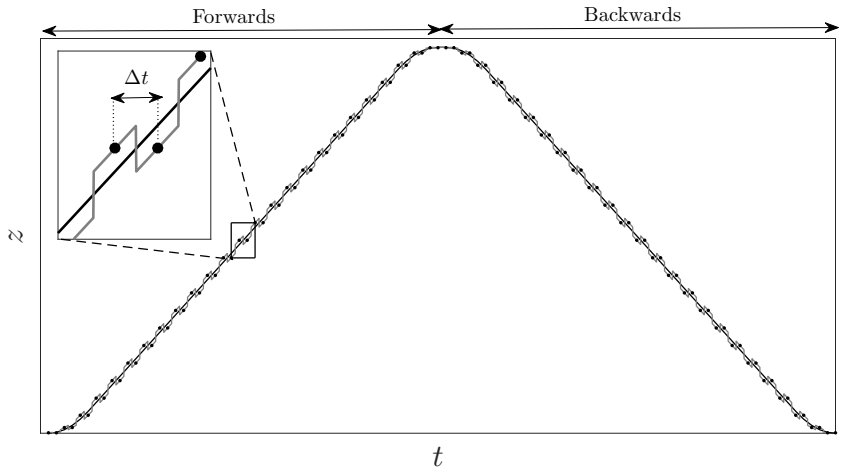}
\caption{Profile of the traverse position with time (black line) and the position of the light sheet due to the displacement of the mirrors (filled circles) that enable pairs of images at discrete spatial locations. The position of the light sheet if it was being fired continuously, rather than pulsed is also shown (grey line)}\label{travPos}
\end{figure*}

Figure \ref{travPos} shows the traverse position (black line) and light sheet position after taking into account the displacement induced by the oscillating mirrors {(filled circles)}. The {grey line} indicates the approximate (due to the finite response time of the mirrors) position of the light sheet if it were a continuous light source. The figure demonstrates how the oscillating mirrors effectively allow the continuous motion of the traverse to be broken down into a series of discrete pairs thus allowing overlapping light sheets for the PIV analysis.   

{To perform the PIV analysis, an accurate calibration procedure is required. This is especially true given the scanned nature of the measurements, as the optical path between the light sheet and the camera varies in time. As the calibration procedure is crucial to the accuracy of the methodology outlined above, a detailed review of the process is given in the following section.}

\subsection{Stereo PIV Calculation}\label{sec:stereoPIVCalc}

To gain quantitative measurements, a calibration is required for the stereo PIV measurements. For the scanning system, accurate calibration mappings are needed at all values of $z$, the scanning direction, within the volume of interest. Moreover, for stereo PIV measurements, gradient information in the $z$ direction is necessary to recover the third component of velocity $w$. In non-scanning systems this gradient information is obtained by either traversing a calibration target (a plane of dots or patterns with known spacing) to several $z$ positions, or by using a multi-plane target where the target has two (or more) planes of dots or patterns at known $z$ locations \citep{prasad2000stereoscopic}. We opt to use a dual-plane target herein, and so, by imaging the calibration target within the experimental set-up, 2D mapping functions $\mathbf{G}_{i}^{k\pm}$ can be calculated. The mappings $\mathbf{G}_{i}^{k\pm}$ map {from} the pixel coordinates of each camera
\begin{equation}
\mathbf{X}_{i}  = \begin{pmatrix}
X_{i} \\
Y_{i} \end{pmatrix} 
\quad\textrm{with}\quad
\mathbf{X}_i\in\mathbb{P}^2,
\end{equation}
to world coordinates 
\begin{equation}
\mathbf{x} = \begin{pmatrix}
x \\
y \\
z_k \pm \frac{\Delta z}{2} \end{pmatrix}
\quad\textrm{with}\quad
\mathbf{x}\in\mathbb{W}^3.
\end{equation}
{Here, $\mathbb{P}^2 \subset \mathbb{R}^2$ is the pixel coordinate space, on the two $z$ planes of the target $z_k \pm \frac{\Delta z}{2}$, where $z_k$ is the mid-plane of the target and $\Delta z$ is the spacing between the two planes, and} where $\mathbb{W}^3 \subset \mathbb{R}^3$ is the world coordinate space. {Specifically,}
\begin{equation}
		\mathbf{x} = \mathbf{G}_{i}^{k\pm}(\mathbf{X}_i),
\end{equation}
{where} $i$ is the camera identifier and $k\pm$ distinguishes between the two planes of the calibration target positioned at $z=z_k \pm \frac{\Delta z}{2}$. {Thus there are two mappings, $\mathbf{G}_{i}^{k+}$ and $\mathbf{G}_{i}^{k-}$, for each camera}. In general, the mapping functions $\mathbf{G}_{i}^{k\pm}$ are not known \emph{a priori} and are commonly determined by least squares fitting of a polynomial function between the known world coordinates $\mathbf{x}$ of the features on the calibration target and the corresponding pixel coordinates of each camera $\mathbf{X}_i$.

For a non-scanning system, with a light sheet centred at $z_k$, the geometric mapping $\mathbf{G}_{i}^{k}$ at $z_k$ can be inferred by linear interpolation between the two mappings at $z_k \pm \frac{\Delta z}{2}$, 
\begin{equation}
\mathbf{G}_{i}^{k}(\mathbf{X}_i) = \frac{1}{2}\left(\mathbf{G}_{i}^{k+}(\mathbf{X}_i) + \mathbf{G}_{i}^{k-}(\mathbf{X}_i)\right).
\end{equation}

To calculate the world displacements $\Delta\mathbf{x}$, we need a 3D mapping. We opt to use an approach similar to \citet{soloff1997distortion} and start by imaging the dual-plane calibration target with both PIV cameras. Then, by identifying a common point on the calibration target, the four pixel coordinates (two from each camera) are used to determine the 3D mapping
\begin{equation}
	\mathbf{x} = \mathbf{F}^k(\mathbf{X}_{\textrm{A}},\mathbf{X}_{\textrm{B}}), 
\end{equation}
where the subscripts A and B distinguish between the two PIV cameras, and $k$ defines the mid-plane location of the calibration target $z_k$ where the mapping is calculated. Given the two planes (separated by $\Delta z$) of the dual-plane target, a polynomial function can be fit, for a given target position $z_k$, with high order dependence in $x$ and $y$ (dependent on the number of features of the target) and linear dependence in $z$. Given the thin light sheet, higher order $z$ dependence is generally not required, but could easily be achieved by including more planes on the multi-plane target or by accurately traversing a planar target in $z$. The world velocities $\mathbf{u}=(u,v,w)$ on a plane positioned at $z=z_k$ can then be found from 
\begin{equation}
	\mathbf{u} \simeq \mathbf{J}_{\textrm{A}}^k\mathbf{U}_{\textrm{A}} + \mathbf{J}_{\textrm{B}}^k\mathbf{U}_{\textrm{B}}, \label{worldDisp}
\end{equation}
where $\mathbf{U}_{i}(\mathbf{X}_i, z_k)=\frac{\Delta \mathbf{X}_i}{\Delta t}$ are the 2D (pixel) velocities for camera $i$ and $\mathbf{J}_{i}^k = \frac{\partial\mathbf{F}^k}{\partial\mathbf{X}_{i}}$ is the $(3 \times 2)$ Jacobian matrix (with units of world/pixel) associated with the mappings at the $z_k$ location of camera $i$.

For the scanning system, the mappings and Jacobian matrices are required for all $z$ in the scan. To this end, the dual-plane target is positioned at a number of $z_k$ locations within the volume of interest. This enables the mappings $\mathbf{G}_{i}^k$ and Jacobians $\mathbf{J}_{i}^k$ (plus inverses of both) to be calculated at several $z$ locations spanning the volume. A least squares fit in $z$ of the polynomial coefficients of the mapping functions is then used to generate $z$-dependent 2D mappings $\mathbf{G_i}(\mathbf{X}_i, z)$ and $z$-dependent Jacobians $\mathbf{J}_{i}(\mathbf{X}_i,z)$ over the whole volume to be scanned. 

The first step in calculating stereo velocities is calculating the 2D velocities in pixel coordinate space $\mathbb{P}^2$ for each of the PIV cameras at a given $z$ location $\mathbf{U}_i(\mathbf{X}_i, z)$ as in eq \eqref{worldDisp}. As the images are initially distorted, largely due to the angular offset of each camera, there are several approaches to calculating the displacements that ultimately need to be on a common grid for the stereo reconstruction of eq \eqref{worldDisp}. The approach adopted herein is to first perform the 2D PIV process for each camera without any image manipulation. This avoids mapping the images to a common grid before the PIV process and eliminates errors due to the image interpolation required. The downside to this approach is that, due to the variable magnification, a constant interrogation window size (in pixels) is not constant in terms of the world coordinates (or particle size). Moreover, for some camera configurations the distortion will be different for each camera, e.g. the left of the image will be magnified for camera A and the right for camera B. The latter of these issues can be avoided, if adequate optical access is feasible, by positioning the two PIV cameras on either side of the light sheet, i.e. both cameras see approximately the same level of magnification across the image. It is worth noting that care has to be taken when positioning the source of the laser sheet using this (or indeed any) approach to ensure that the scattering of the light from the seeding particles is comparable for both cameras \citep{willert1997stereoscopic}. The velocities $\mathbf{U}_i\left(\mathbf{X}_i, z\right)$ in pixel coordinate space $\mathbb{P}^2$ at a known $z$ location are mapped to a common grid in world coordinate space $\mathbb{W}^3$. The $z$ position of the velocity fields is used so that the pixel velocities of the two cameras in pixel coordinate space $\mathbb{P}^2$ can be mapped to the corresponding world coordinate space $\mathbb{W}^3$ using the $z$ dependent 2D mappings generated during the calibration
\begin{equation}
{\mathbf{U}_i\left({\mathbf{X}_i}, z\right)} \xmapsto{\quad \mathbf{G}_i(\mathbf{X}_i, z) \quad} {\hat{\mathbf{U}}_i(\mathbf{x})}.
\end{equation}
The Jacobian $\mathbf{J}_{i}$, initially in $\mathbb{P}^2$ space, is also mapped to world coordinate space $\mathbb{W}^3$ in the same manner
\begin{equation}
\mathbf{J}_i(\mathbf{X}_i, z)\xmapsto{\quad\mathbf{G}_i(\mathbf{X}_i, z)\quad} \hat{\mathbf{J}}_i(\mathbf{x}).
\end{equation}
Note that the values of $\hat{\mathbf{J}}_i$ still represent the world units per pixel. From $\hat{\mathbf{U}}_\textrm{A}$ and $\hat{\mathbf{U}}_\textrm{B}$, the two sets of pixel velocities now in $\mathbb{W}^3$, we can calculate the velocity field in world units and in $\mathbb{W}^3$ using
\begin{equation}
\hat{\mathbf{u}} \simeq \hat{\mathbf{J}}_\textrm{A}\hat{\mathbf{U}}_\textrm{A} + \hat{\mathbf{J}}_\textrm{B}\hat{\mathbf{U}}_\textrm{B}.\label{finalVel}
\end{equation}
We do this for each pair of 2D2C planes produced by the 2D PIV algorithm for cameras A and B to construct a sequence of 2D3C planes at different $z$ locations in the scan. These 2D3C planes are finally combined to construct volumetric 3D3C measurements. 

\subsection{Error}

For each $z$ location of the scan, an estimate of the error in the stereo reconstruction can be determined by back-projecting the world velocities $\hat{\mathbf{u}}$ onto the velocities (in pixels/frame) of the two cameras. For convenience, we do this calculation on the common world grid to obtain $\hat{\mathbf{U}}_{i}^*$ as follows
\begin{equation}
\begin{pmatrix}
\hat{\mathbf{U}}_\textrm{A}^* \\
\hat{\mathbf{U}}_\textrm{B}^* \end{pmatrix}  = \hat{\mathbf{J}}^{-1}\hat{\mathbf{u}}\label{UABackProj}
\end{equation}
where $\hat{\mathbf{J}}^{-1}(\mathbf{x})$ is the inverse Jacobian in world coordinate space $\mathbb{W}^3$ found from
\begin{equation}
\mathbf{J}^{-1}(\mathbf{X}_i, z)\xmapsto{\quad\mathbf{G}_i(\mathbf{X}_i, z)\quad} \hat{\mathbf{J}}^{-1}(\mathbf{x}),
\end{equation}
where $\hat{\mathbf{J}}^{-1} = \frac{\partial\mathbf{X_i}}{\partial\mathbf{F}_{i}^{-1}}$ is the $(4\times 3)$ Jacobian matrix (with units of pixel/world) of the inverse 3D mapping $\mathbf{F}_i^{-1}$. Note that $\mathbf{F}_i^{-1}(\mathbf{x})$ is our estimate of the projection function $\mathbf{P(\mathbf{x}})_i$ given in eq \eqref{2DProject} and is calculated using a least squares approach {similar to that} for the forward 3D mapping $\mathbf{F}(\mathbf{X}_a, \mathbf{X}_b)$. A measure of the error can be obtained by comparing the back-projected velocities with the velocities calculated by the PIV algorithm mapped to $\mathbb{W}^3$. Assuming equal weighting among each of the velocity components, we construct the error field {from the magnitude of the vector difference}
\begin{equation}
E(\mathbf{x}) = \frac{1}{4}\left(\lvert\lvert\hat{\mathbf{U}}_\textrm{A}^* - \hat{\mathbf{U}}_\textrm{A}\rvert\rvert + {\lvert\lvert\hat{\mathbf{U}}_\textrm{B}^* - \hat{\mathbf{U}}_\textrm{B}\rvert\rvert}\right).\label{BackProjErr}
\end{equation}
This error estimate can be used as an additional quality check of the velocity fields by removing vectors where the reconstruction error in ${E}(\mathbf{x})$ exceeds some threshold value (typically chosen to be 0.5 pixels/frame as in \cite{wieneke2005stereo}).  

\subsubsection{Calibration Refinement}\label{sec:refinement}

An important step in calculating the velocities using stereo PIV is the calibration refinement technique. The stereo reconstruction of velocity, as outlined in \S\ref{sec:stereoPIVCalc}, relies on the calibration between each camera and the world coordinates being accurate. This ensures that a velocity perceived from one camera is correctly reconstructed with a velocity from the other camera, i.e. the two velocities are coincident in world coordinates. Although, in practice, the calibration procedure is performed carefully, it is difficult to perfectly align the calibration target with the position of the light sheet and even small discrepancies can yield large errors in the stereo reconstruction \citep{wieneke2005stereo}. {With the scanning system this is complicated further as any small misalignment between the laser beam and the optics on the traverse carriage can lead to a systematic positional error in the light sheet.} To correct for this, we opt to use an iterative two-step calibration refinement technique using a methodology similar to \citet{willert1997stereoscopic}. The method makes use of the simultaneously acquired raw particle images from each of the PIV cameras that are then mapped, using their individual coordinate mapping $\mathbf{G}_i(\mathbf{X}, z)$, to world coordinates. If the light sheet and the calibration target were perfectly coincident, and the light sheet was infinitesimally thick, the two particle images should perfectly overlay each other. To determine how well the light sheet and coordinate system are aligned, a disparity map is calculated by cross-correlating the two images. 

The first step of the refinement technique is to refine the $z$ position of the light sheet. On the first iteration, the approximate $z$ position is known by initiating the traverse of the light sheet from a given location, e.g. positioned so the light sheet just intersects a boundary of the apparatus, and by knowing the subsequent displacement of the traverse along with the superimposed displacements due to the oscillating mirrors. This approximate location is known for all $2N$ frames of the scan sequence, where $2N$ is the total number of frames in a complete scan sequence ($N$ forwards and $N$ backwards), $\mathbf{z}_l^{n-1}=(z^{n-1}_0,z^{n-1}_1,...,z^{n-1}_{2N-1})$ where $n$ is the current iteration. For each frame $j$, the approximate $z^{n-1}_j$ location of the light sheet is used in the 2D mappings $\mathbf{G}_i(\mathbf{X}_i,z^{n-1}_j)$ to map the images to world coordinate space $\mathbb{W}^3$. The pair of images are cross-correlated, over the whole of the overlapping region between the two images, and then a bisection method used to minimise the disparity between the images by adjusting $z^n_j$ in $\mathbf{G}_i(\mathbf{X}_i,z^{n}_j)$, i.e. the $z$ location of the coordinate mappings is changed until the cross-correlation peak is a maximum with zero shift between the images. The output of the bisection algorithm is monitored to check that the optimal light sheet position is within given bounds of the initial guess $\mathbf{z}_l^{n-1}$, typically $\pm \delta z$ with these bounds depending on the certainty of the initial light sheet position. Before performing the cross-correlation, the images are preconditioned by removing the background. This is accomplished by removing a low-order fit to the image, but other background-removal strategies should work equally well (e.g. a background image found via low pass filtering of the images, etc.). The intensities are also normalised between the two image streams so that each has approximately the same range in intensity. To further improve the quality of the peak in the cross-correlations, the cross-correlations are averaged over a suitable number ($\sim 5$) of scan periods (i.e. the same traverse position but a different image pair). Doing this for every frame in the scan, a refined light sheet position $\mathbf{z}_l^n$ is determined. To make the refined light sheet position more robust to any outliers, the mean offset $\overline{z_l}=\overline{\mathbf{z}_l^n-\mathbf{z}_l^{n-1}}$ between $\mathbf{z}_l^n$ and $\mathbf{z}_l^{n-1}$ can be found and the updated light sheet position given by $\mathbf{z}_l^{n}=\mathbf{z}_l^{n-1}+\overline{z_l}$. This inherently assumes that the difference between the two light sheet positions is a constant which is usually adequate (the variance of $\mathbf{z}_l^n-\mathbf{z}_l^{n-1}$ is typically $O(10^{-2})$ mm$^2$). However, a more general approach would be to use a spline approximation to $\mathbf{z}_l^n$ with knots positioned at the beginning and end of the constant velocity section of the scanning. Using the spline method, one would then account for the displacement of the light sheet by the oscillating mirrors using a moving average over a pair of light sheet positions, i.e. $z_0^n = z_1^n = \frac{1}{2}(z_0^n + z_1^n)$, $z_2^n = z_3^n = \frac{1}{2}(z_2^n + z_3^n)$, etc.

The second step of the refinement allows for finer warping of the coordinate mappings to correct for any residual misalignment between the light sheet position and the calibration target used to calculate the mappings (e.g. small rotations of the target relative to the light sheet). This is achieved by again mapping pairs of raw particle images (one from each camera) to world coordinates (using the newly refined light sheet positions $\mathbf{z}_l^n$) and subdividing the resultant pair into small interrogation windows. These interrogation windows are then cross-correlated, effectively running a PIV algorithm on the image pairs. Any mismatch is detected by the algorithm and used to determine the optimal shift to realign the particle images from the two cameras. To remove any unwanted outliers from the disparity map, a least squares approach is used to fit a polynomial to the disparity map over the plane $(x, y)$ and the scanning direction $z$. The polynomial fit to the disparity map is applied to each of the 2D velocity fields and the Jacobians (at the appropriate $z$ location), with $+\frac{1}{2}$ shift for camera A and $-\frac{1}{2}$ shift for camera B, before the the world coordinate velocities $\hat{\mathbf{u}}$ are calculated using eq \eqref{finalVel}.

{The two-step approach can then be iterated until $|\mathbf{z}_l^n - \mathbf{z}_l^{n-1}|<< \delta z$,} updating the refined light sheet position as well as warping the particle images according to the disparity map calculated in the $n$th iteration of the second step. In practice, a single iteration of the two-step algorithm is enough to {satisfy the requirement that} any residual correction is much less than the thickness of the light sheet.

\subsection{PLIF Calculation}\label{Sec:LIFCalc}

To determine the density field, the signal from the fluorescent dye added to the flow has to be calibrated. In general, light passing through the fluid is going to be attenuated as it travels through the various media, including the water, the PIV particles, and the fluorescent dye itself. However, in a non-reactive medium and for low concentrations of fluorescent dye (and particles), as is often the case when used in conjunction with a high power pulsed laser, the attenuation is negligible. If the camera response is linear then
\begin{equation}
	{I} = I_0 + I_{laser}\alpha {C},
\end{equation}
where ${I}$ is the intensity perceived by the camera, $I_0$ is a background reference image that contains any static noise from the camera along with any externally-induced light field, $I_{laser}$ is the intensity of the laser, {${C}$ is the concentration of dye and $\alpha$ is a constant} (see \citet{crimaldi2008planar} for details). The calibration of the experimental images then simply reduces to
\begin{equation}
{C}(\mathbf{X},z) = \frac{{I}(\mathbf{X},z) - {I}_0(\mathbf{X},z)}{{I}_1(\mathbf{X},z) - {I}_0(\mathbf{X},z)}C_1,\label{LIFCalc}
\end{equation}
where ${C}(\mathbf{X},z)$ is the calculated concentration field, ${I}_1(\mathbf{X},z)$ is a reference image containing a known homogeneous mixture of dye that can be used to remove any spatial variation present in the light sheet and establish the concentration of dye in the images, and $C_1$ is the dye concentration for which $I(\mathbf{X},z)$ = $I_1(\mathbf{X},z)$. These  two reference images are determined as follows. First, a scan-position-dependent background reference image ${I}_0(\mathbf{X},z)$ is calculated at each $z$ position of the scan by recording $O(10)$ scans of the volume containing no fluorescent dye and taking the mean at each $z_l$ location. To minimise errors, no fitting is used for the background images and a distinct image is found for every $z_l$ location separately for the forward and backward scans (i.e. there are $2N$ separate background images in total). In a similar manner, a second reference image ${I}_1(\mathbf{X},z)$ is calculated by averaging over $O(10)$ scans of the volume containing a homogeneous mixture of the highest concentration of dye in the experiment $C_1$, again producing $2N$ images.

Finally, the concentration measurements found from eq \eqref{LIFCalc} are mapped to world coordinate space $\mathbb{W}^3$. As discussed in \S \ref{sec:stereoPIVCalc}, this is achieved in the scanning system using a least squares mapping $\mathbf{G}_C(\mathbf{X}_{\textrm{C}}, z)$ calculated from the images recorded of the calibration target for the PLIF camera (camera C). The concentration $\hat{C}$ in world coordinate space $\mathbb{W}^3$ can then be found from
\begin{equation}
{C}(\mathbf{X},z)  \xmapsto{\quad \mathbf{G}_{\textrm{C}}(\mathbf{X}_{\textrm{C}}, z) \quad} \hat{{C}}(\mathbf{x}).
\end{equation}

As for the PIV cameras (A and B) the coordinate system of camera C can be refined using the strategy discussed in \S \ref{sec:refinement}. In this case, only the second step of refinement is needed, and the coordinate system of camera C can be refined using simultaneously acquired particle images from camera C, with its longpass filter removed, and one (or both) of the PIV cameras. To ensure all of the coordinates systems lie in the same plane, the disparity map is calculated using the refined coordinate system of camera A or B (or both) with the full disparity map (or the average disparity map if the refinement is conducted with both PIV cameras) applied to camera C images only.

To relate the concentration to the the density field $\rho$ we have
\begin{equation}
\rho =  \frac{\Delta\rho}{C_1}\hat{{C}} + \rho_{min},\label{LIFToWorld}
\end{equation}
where $\Delta\rho=\rho_{max} - \rho_{min}$ and $\rho_{max}$, $\rho_{min}$ are the densities corresponding to the fluid with the maximum concentration of dye and the fluid containing no dye, respectively.

It is worth noting that, in the scanning system, we have twice as many $\rho$ fields as velocity fields because the PLIF calculation only requires a single image (compared to the two required to calculate a velocity field using PIV). That being said, there is a choice in how to make use of the extra information. We could use either a single $\rho$ field, corresponding to the first or second raw particle image used in the PIV calculation, or the mean of the two $\rho$ fields could be taken to help remove random noise from the data. Alternatively, the velocity information calculated for corresponding pair of $\rho$ fields could be used to advect the fields forward and backward in time by $\pm\frac{1}{2}{\mathbf{u}}{\Delta t}$ before taking the mean of the two fields.

\section{Experimental Set-up}\label{expSetup}

The methodology outlined in this paper was used to perform volumetric measurements of the buoyancy-driven exchange flow in an inclined duct similar to the flow studied by \citet{meyer2014stratified} and \citet{lefauve2018}. We opted to use this particular experiment with our new measurement method as it was stratified, and so knowledge of the density field was desired, and naturally has 3D spatial structure due to the confining duct boundaries. Moreover, turbulent flow states are possible that result in a complex 3D structure to both the velocity field and density field. To this end, we chose our parameters such that we were in the `Turbulent Regime' identified by \citet{meyer2014stratified}.

\subsection{Experimental Apparatus}

\begin{figure*}[h!]
\centering
\includegraphics[width =  0.95\textwidth]{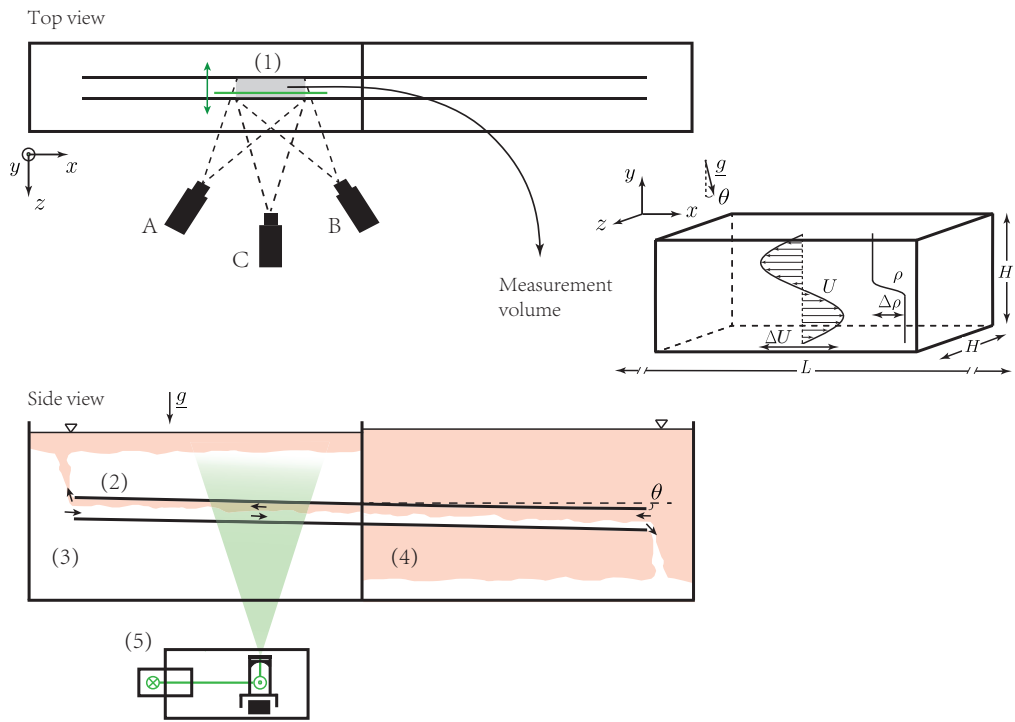}
\caption{{Sketch of experimental set-up. Top view: (1) approximate region of the duct where the volumetric measurements are made by the three cameras (A and B for stereo PIV and C for PLIF). Side view: (2) inclined duct that confines the exchange flow between the two reservoirs; (3) reservoir containing the relatively dense fluid; (4) reservoir containing the relatively light fluid (tagged with Rhodamine 6G for PLIF). The laser beam was emitted from the scanning system (5) and illuminated the flow through the base of the duct. Measurement volume: A 3D view of the measurement volume (1) is shown clarifying the orientation of the coordinate system used}}\label{SIDSchem}
\end{figure*}

The experimental set-up is shown in figure \ref{SIDSchem} and consisted of two reservoirs joined only through a  duct. Each reservoir had dimensions $100 \times 20 \times 50$ cm and the duct had a square cross-section of height and width $H=4.5$ cm with a total length $L=135$ cm. The duct and the reservoirs were made of Perspex (acrylic) and were of good optical quality. The duct passed through a flexible gasket that was located in the central wall separating the two reservoirs. This allowed the duct to be tilted whilst still maintaining a seal between the reservoirs. The reservoirs were filled with different density fluids $\rho_0\pm\Delta\rho/2$ where $\Delta\rho=0.0117$ g/cm$^3$. The duct, that joined the two reservoirs, was tilted at an angle $\theta=5^\circ$ from the horizontal, see figure \ref{SIDSchem}. 

We chose to orientate the $x$ axis with the streamwise direction of the flow, along the duct, with $z$ in the spanwise direction (the scanning direction in this set-up). The $y$ axis was then inclined at an angle of $5^\circ$ from the vertical upwards direction. Note that, rather than using the convention of having the $z$ axis aligned with gravity {as is common} in the stratified flows literature, {our chosen} orientation {matches that} used in the methodology of this paper (i.e. $z$ in the scanning direction). All coordinates had their origin in the middle of the duct, such that $-L/2 \leq x \leq L/2$ and $-H/2 \leq y,z \leq H/2$. Our measurement volume was approximately $6H$ in the streamwise direction and spanned the full vertical and spanwise extents of the duct (both $H$ here). The measurement volume was offset from the centre of the duct (in the relatively dense layer, as shown in figure \ref{SIDSchem}) to avoid the flexible gasket that separated the two reservoirs.

\subsection{Optical Components}

For the experiments presented in this paper, a frequency-doubled dual-{cavity} Litron Nano L100 Nd:YAG laser was used as the light source, with a wavelength of 532 nm. The laser had a maximum repetition rate $f_{laser}=100$ Hz for each of the two laser cavities. For the galvanometer mirrors we used two Thorlabs GVS311/M units attached to a Thorlabs GPS011 driver/power supply. The mirror orientations were controlled by an analogue signal that determined the subsequent rotation of the galvanometer. Given that only small rotations of the mirror ($< 1^\circ$) were needed in the current system, the highest resolution setting for the controller was used (a $\pm 10$ V input signal corresponded to a $\pm 10^\circ$ rotation). The galvanometers (with the mirrors attached) had an angular resolution of ${8\times 10^{-4}}^\circ$ and both mirrors had an aperture  $A_m = 10$ mm. In the arrangement here, the mirrors were mounted so that their resting state was $\theta_m = 9.9^\circ$ with a horizontal and vertical distance between the centres of $L_c = 18.8$ cm and $H_c = 6.8$ cm, respectively. Given this geometric arrangement, the aperture of the mirrors, and the diameter of the beam from the laser ($\sim 4$ mm), the beam could be displaced by $\pm 2.9$ mm {(with a relative rotation of $\pm 0.4^\circ$ to the mirrors resting angle $\theta_m$)}, which gives a maximum displacement between image pairs of $D_m=5.8$ mm. Note that this maximum allowable beam displacement $D_m$ (see figure \ref{LaserMirrors}) set an upper limit on the speed of traversing $V$ for a fixed inter frame time $\Delta t$ as, to get overlapping light sheets, we require $V<\frac{D_m}{\Delta t}$. For the current system, in single-pulse mode, this meant that the oscillating mirrors limited the the maximum speed of the traverse to $V|_{max}= 58$ cm/s. Note that, in practice, the limit was less than this due to the limited aperture of the sheet-producing optics and the non-negligible divergence of the laser source.

To produce a scanning light sheet from the nominally circular beam emitted from the laser, a system of cylindrical lenses was mounted on a traversing carriage\footnote{Note that for historical reasons a Dantec focusing module (9080X0911) was attached to the cylindrical lenses in the current arrangement. Due to the spherical nature of the focusing lenses, {a simple parallel displacement of the beam would not} provide parallel light sheets. {Instead}, the rotation of the individual mirrors was fine tuned so that the laser entered the optics at a slight angle thus allowing nearly parallel light sheets with good overlap between image pairs. In an ideal system, the focusing module would be placed before the oscillating mirrors.}. The traverse carriage was mounted on an Igus DryLin SAW rail system. The system was chosen due to its plastic bearings, avoiding metal on metal contact introducing vibrations into the system, and as it was impervious to salt/water. It was also chosen for its relatively wide separation between the rails that minimised any roll of the carriage during motion. To move the carriage, a stepper motor was chosen for the simplicity of positional control without relying on a separate position resolver (as, for example, would be the case with a conventional servo motor). The traverse carriage was attached to a stainless steel lead screw of pitch 2 mm, diameter 10 mm, and with a maximum travel of 500 mm. Due to the inertia of the carriage and friction from the rails, the carriage could not simply be put into motion at a constant velocity and so a constant acceleration/deceleration phase was implemented. For the experimental results shown here, the total time spent accelerating/decelerating was $\sim 15\%$ of the scanning period. As a consequence, the spacing of measurements in $z$ was non-uniform at the beginning and end of the scan.

Inevitably, vibrations were introduced into the traverse system. However, the vibrations introduced into the carriage because of the stepper motor were found to be negligible in practice (micro stepping was used to help minimise this). Moreover, to check if the traverse system deteriorated over time, {a three-axis accelerometer was} housed on the traverse carriage to monitor the vibration levels during the scanning. Even after multiple uses in the lab environment, the accelerometer data showed that the vibration levels of the carriage had not increased. To minimise any residual vibrational feedback from the complete traversing system onto the experiment itself, the traversing system and laser were mounted on a large (60 cm x 120 cm) honeycomb optical breadboard (Thorlabs PBG51507), kept separate from the bench on which the experiment was mounted.

To image the flow, three Teledyne Dalsa Falcon2 8M cameras were used. {These CMOS ten-tap CameraLink} cameras had a maximum resolution of 3320 x 2502 pixels, however, given the aspect ratio of the duct, a reduced resolution for each of the cameras (3320 x 824 pixels) was used. The three cameras (two for PIV and one for PLIF) were fixed in position to one side of the duct (as shown in figure \ref{SIDSchem}), and the angular offset between the two PIV cameras was chosen to be $\sim 80^{\circ}$. To improve focusing across the image, both PIV cameras were fitted with Scheimpflug adapters. It was not deemed necessary to install liquid filled prisms in between the stereo PIV cameras and the volume of interest but we note that in some situations this could be advantageous, as discussed by \cite{prasad1995scheimpflug}. The PLIF camera had its optical axis normal to the light sheet and so did not require a Scheimpflug adapter. The PIV cameras were fitted with Micro Nikkor 60 mm f/2.8D lenses at aperture f/8 and the PLIF camera had an Nikkor 50 mm f/1.2D lens at aperture f/1.2 (so that only a small concentration of fluorescent dye was required in the dyed layer, consistent with the linear relation assumed in the calibration calculation, see \S \ref{Sec:LIFCalc}). So that each camera had an adequate depth of field, the PIV cameras were located $\sim$0.6 m away from the measurement volume with the PLIF camera further back (because of the increased aperture) at $\sim$1 m. This distance was chosen in-situ and enabled both the particle images and the dye images to be adequately focused across the whole depth of the volume. As it was not feasible to only seed particles in the region of interest, a transparent Perspex (acrylic) box filled with water (but with no particles) was positioned between the inner wall of the reservoir and the outer boundary of the duct itself in the optical path between the cameras and the light sheet. This `optical box' removed a large portion of noise present in the raw images for all cameras {that would otherwise occur} due to the optical path between the light sheet and cameras encountering a large number of particles when high seeding densities were used. 

\subsection{Experimental Protocol}

Prior to the experiment, the system was calibrated as discussed in \S \ref{sec:stereoPIVCalc}. The calibration refinement step was typically performed with particle images captured during the experiment with a second set of refinement images captured between Camera C (the PLIF camera with the filter removed) and Camera A either before or after the experiment. After calibrating, both of the reservoirs were filled with fresh water and allowed to settle overnight to help degas the fluid, to minimise bubble formation during the experiment, and reach the ambient temperature. The duct was then inclined to the desired angle (as determined by a Digi-Pas DWL-280Pro digital inclinometer) and the end open to the reservoir that would contain the denser fluid was temporarily sealed (to avoid unwanted exchange of fluid between the reservoirs before the start of the experiment). 

The density difference between the reservoirs was achieved using two salt solutions, NaNO$_3$ and NaCl, such that the refractive indices of the solutions were matched at 532 nm (the wavelength of the laser source). This particular combination of salts was used as they, to a good approximation, mix linearly and have similar diffusivities at the low concentrations required here \citep{olsthoorn2017three}. These salt solutions were added to the appropriate reservoirs to establish the desired density difference. The densities of the fluids in the two reservoirs were measured at a temperature of $20~ ^\circ$C (the same temperature as the laboratory environment) using an Anton Paar DMA 5000 density meter. To ensure accurate PIV and PLIF, the refractive indices of the two solutions were matched with a relative error of $\Delta n/ n \approx 10^{-4}$ and verified by a handheld refractometer (Reichert Technologies Goldberg) using a light source with a green filter to approximate the wavelength of the laser.

For PIV, polyamide particles with a diameter of $50$ $\mu$m and nominal density between $1.02-1.03$ g/cm$^3$ were used to seed the flow. The particles were chosen due to their small ratio of settling to mean flow velocities $V_p/(\sqrt{g'H}) = 4.17 \times 10^{-4}$ while still being large enough to provide a clear particle image.The polyamide particles were added to the flow with a small amount ($\sim 5$ ml) of Magnum$^{\textrm{\tiny{\texttrademark}}}$ Rinse Aid to prevent aggregation of particles. Enough seeding particles were added to the flow to ensure that that there were always $\gtrsim 5$ particles in any interrogation window.

At this stage, the first sequence of calibration images $I_0$ for the PLIF measurement could be captured as discussed in \S \ref{Sec:LIFCalc}. Note that before recording any calibration images for the PLIF measurements, or beginning capture during the experiment, the laser and associated optical components were allowed to warm up for at least 30 s. This avoided a non-trivial time-varying {change} in spatial structure {and} magnitude {of the light sheet. Such changes would otherwise hamper the PLIF calculation.} After this first sequence of calibration images had been captured, rhodamine 6G was added to the reservoir containing the relatively light NaCl solution and mixed giving a final concentration of rhodamine 6G in the reservoir of $C_1 = 15$ $\mu$g/L. Adding the dye to this reservoir {avoided having the light sheet} pass through a layer of dyed fluid before entering the measurement volume. After ensuring the dyed fluid was mixed into the reservoir and the duct, the second sequence of PLIF calibration images $I_1$ were captured. 

The experiment could then be {started} by simply removing the temporary seal from the end of the duct. Initially, there was a transient stage of the experiment as a gravity current of denser un-dyed NaNO$_3$ fluid propagated through the duct into the opposite reservoir. After this period of transient flow had finished we began taking the three-dimensional measurements reported here.  

\subsection{Measurement Resolution and Processing}\label{measProc}

The resolution of the measurements for the set-up discussed here were primarily set by the laser system and the choice of camera. The resolution in the scanning direction was predominantly set by the laser as the maximum number of velocity fields that could be obtained in 1 s was 100 (achieved by firing the two cavities of the laser out of phase at their maximum repetition rate 100 Hz). Moreover, the precise choice of laser and optical components used to produce the light sheet controlled the thickness of each plane $\delta z$, and the subvolume over which each of the initial 2D pixel displacements were determined. For the measurements herein, given the small depth required to image the duct (45 mm) and the typical thickness of the light sheet in the current set-up (1-2 mm), we chose to discretise the domain into $\sim 40$ planes. This avoided unnecessary overlapping of light sheets given that the uncertainty in the position of the velocity fields in the $z$ direction will be the same order as the light sheet thickness. {To adequately resolve the particle displacements, the system was in double-pulse mode, with a spacing between laser pulses $\Delta t = 7$ ms and camera frame rate $100$ fps.}   

The resolution of the in-plane measurements in ($x$, $y$) was set by the resolution of the camera, the seeding density of the PIV particles, and the PIV algorithm used. For the results shown here, all of the raw images were processed using DigiFlow \citep{dalziel2007simultaneous}. The processing used the DigiFlow 2017a PIV algorithm, selecting an {initial} interrogation window of height and width 31 pixels and a spacing of 12 pixels (both horizontally and vertically) equivalent to a 60\% overlap of interrogation windows. The algorithm underlying these PIV calculations has some important differences from most PIV implementations. First, as introduced for synthetic schlieren by \citet{dalziel2000whole}, the pattern matching kernel is based on an $L_1$ norm measure of the differences between images in the interrogation windows and is used in place of the more common $L_2$ norm of a cross-correlation function. This kernel is applied iteratively, utilising displacement information from previous passes to ‘advect’ the image pair captured at $t = t_i \pm \Delta t/2$ to their anticipated state at $t = t_i$. {The {size, shape, and weighting profile} of each interrogation window are} adjusted (based on their information content and the spatial gradients in the displacement field) during this process to faithfully capture high gradients while achieving a low level of noise. Additionally, a strategy of weighting or removing anomalous pixels increases further the robustness of the results to particles entering or leaving the light sheet.

The effective resolution is not constant across the field of view due to the distortion across the images, caused by the angles of the cameras relative to the light sheet. The resolution also varies as the light sheet is scanned, a consequence of the cameras being stationary, so there is higher resolution when the light sheet is closest to the cameras. However, with the set-up detailed here, the loss of resolution because of the scanning was negligible (a change in pixels of $<5\%$ for a given physical length over the depth of the scan) compared to the distortion of the images (a change of $\sim20\%$ across the image). Therefore, the maximum resolution of the velocity measurements in the $(x,y)$ plane reported here was 1 velocity vector every 0.51 mm and the minimum resolution was 1 velocity vector every 0.63 mm. The total yield of vectors in a scan, acquired every 0.77 s, was approximately $400 \times 100 \times 40$ $(x,y,z)$. Indicative of the quality of the vector fields, the reconstruction error field $E(\mathbf{x})$ (see eq \eqref{BackProjErr}) {had an average in the $(x,y)$ plane of $\langle{E}\rangle_{x,y}<6.7 \times 10^{-2}$ (spanning all $z$ in the scan) with a maximum standard deviation $\sigma_{max} = 4.6 \times 10^{-2}$, both in units of pixel/frame.} It was observed that the error systemically increased in $z$ or increasing distance from the cameras. However, it is worth noting that, due to size constraints of the current dual-plane calibration target, the calibration could only be performed at three distinct $z_k$ locations with $z_k \leq 0$ (so we could only calibrate over half of the spanwise extent of the duct situated closest to the cameras). This is a possible cause of the increasing trend observed as a similar trend is also found if we simply go between the forward and inverse mappings (i.e. use the forward mapping on the back-projected fields $\hat{\mathbf{U}}_{i}^*$ to obtain $\hat{\mathbf{u}}_{i}^*$ and then the inverse mappings on this field to obtain a second set of back-projected mappings that can be compared to $\hat{\mathbf{U}}_{i}^*$). 

The density fields in world coordinates were calculated from from eq \eqref{LIFCalc} and eq \eqref{LIFToWorld} using the raw images and the two calibration images $I_0$ and $I_1$. Note that, for the data shown here, only the first of the two PLIF images was used to reconstruct the density field. As no interrogation window is required for the PLIF calculation, the resolution of the PLIF measurements is higher than those of the velocities. In the current set-up, there was a density measurement in the $(x,y)$ plane every 0.1 mm. In total, there were approximately $3000 \times 500 \times 40$ $(x,y,z)$ density measurements every 0.77 s, given by the resolution of the camera and the number of frames in a scan. For both the velocity and density fields, the resolution in the scanning direction $z$ is $1.26$ mm with an accuracy set by the light sheet thickness $\delta z$ (1-2 mm). 

Before presenting the data, the density fields were processed to remove line artifacts from $(x,y)$ planes (due to light rays passing through residual air bubbles, clusters of particles, and imperfections in the tank surfaces) using a three-step approach. As all of the rays emanated from some virtual origin below the tank and spread with the light sheet, the first step straightened the lines by mapping the $(x,y)$ planes into a `ray coordinate' system producing planes with vertical line artifacts. The second step removed these vertical line artifacts from the planes using a wavelet method described by \cite{munch2009stripe}. The planes, now without line artifacts, were mapped back to world coordinates. Finally, after removing the line artifacts, the density data was interpolated onto the lower resolution grid of the velocity data after being median filtered over a suitably sized window.

\section{Experimental Results}\label{results}

For the results presented here, we choose to non-dimensionalise velocities $\mathbf{u}$ by $\sqrt{g'H}$, where $g'=g\Delta\rho/\rho_0$ is the reduced gravity, and all lengths by $H/2$. As a result, {with the duct angle fixed at $\theta=5^\circ$}, we can construct a Reynolds number $Re=\sqrt{g'H}H/(2\nu)=1516$ to characterise the flow. The natural time scale to non-dimensionalise time is the advective time $({1}/{2})\sqrt{H/g'}$. Finally, the non-dimensional density field is $\tilde{\rho} = 2(\rho - \rho_0)/(\Delta \rho)$ such that $-1 \le \tilde{\rho} \le 1$. Therefore, our final measurement volume in non-dimensional units was $17.4 \leq \tilde{x} \leq -6.3$, $-1 \leq \tilde{y} \leq 1$, $-1 \leq \tilde{z} \leq 1$ and a total of $561$ advective time units were captured with a non-dimensional time between volumes of $2.40$. The tilde of non-dimensional variables will be dropped henceforth.

\begin{figure*}[h]
\centering
\includegraphics[width = 0.95\textwidth]{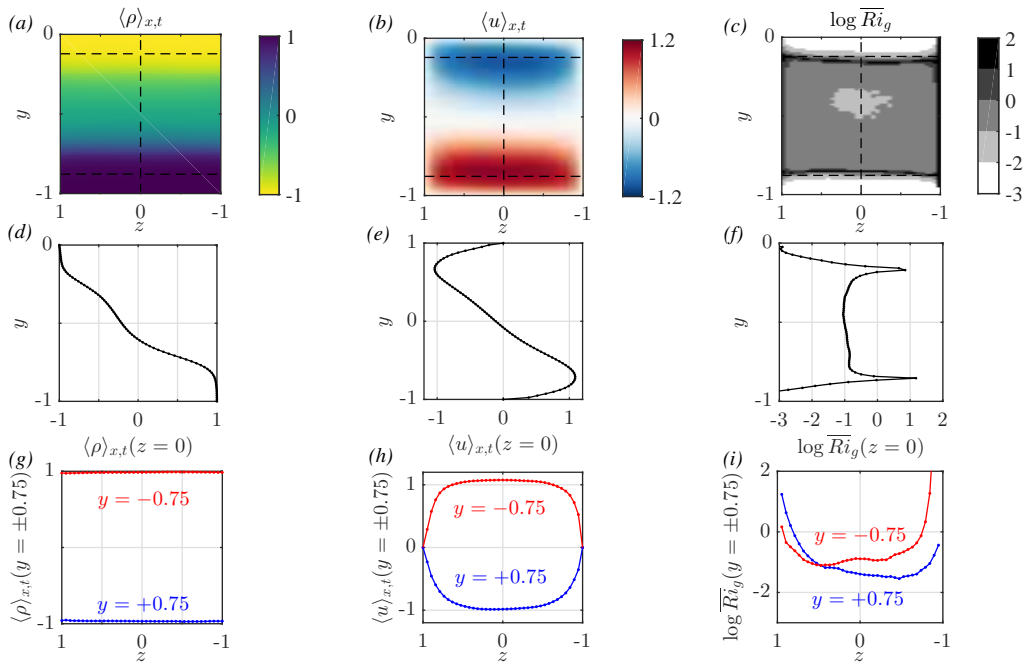}
\caption{Means from the volumetric data. Each column corresponds to a different quantity: $\rho$, $u$ (streamwise velocity), and $\overline{Ri}_g$. The first row (a)-(c) shows the $\langle \cdot \rangle_{x,t}$ averaging (or combination of averages in (c) to form $\overline{Ri}_g$) of each quantity (averaged in $x$ and $t$) on a $(y,z)$ slice. The second row (d)-(f) shows a single vertical profile of the same data at the centre of the duct ($z=0$) that corresponds to the dashed line shown in (a)-(c). The final row (g)-(i) shows the variation in the spanwise direction $z$ of the same data for two different vertical locations, again shown by the dashed lines in (a)-(c)}\label{fig:means}
\end{figure*}

\begin{figure*}[!hp]
\centering
\includegraphics[width = 0.95\textwidth]{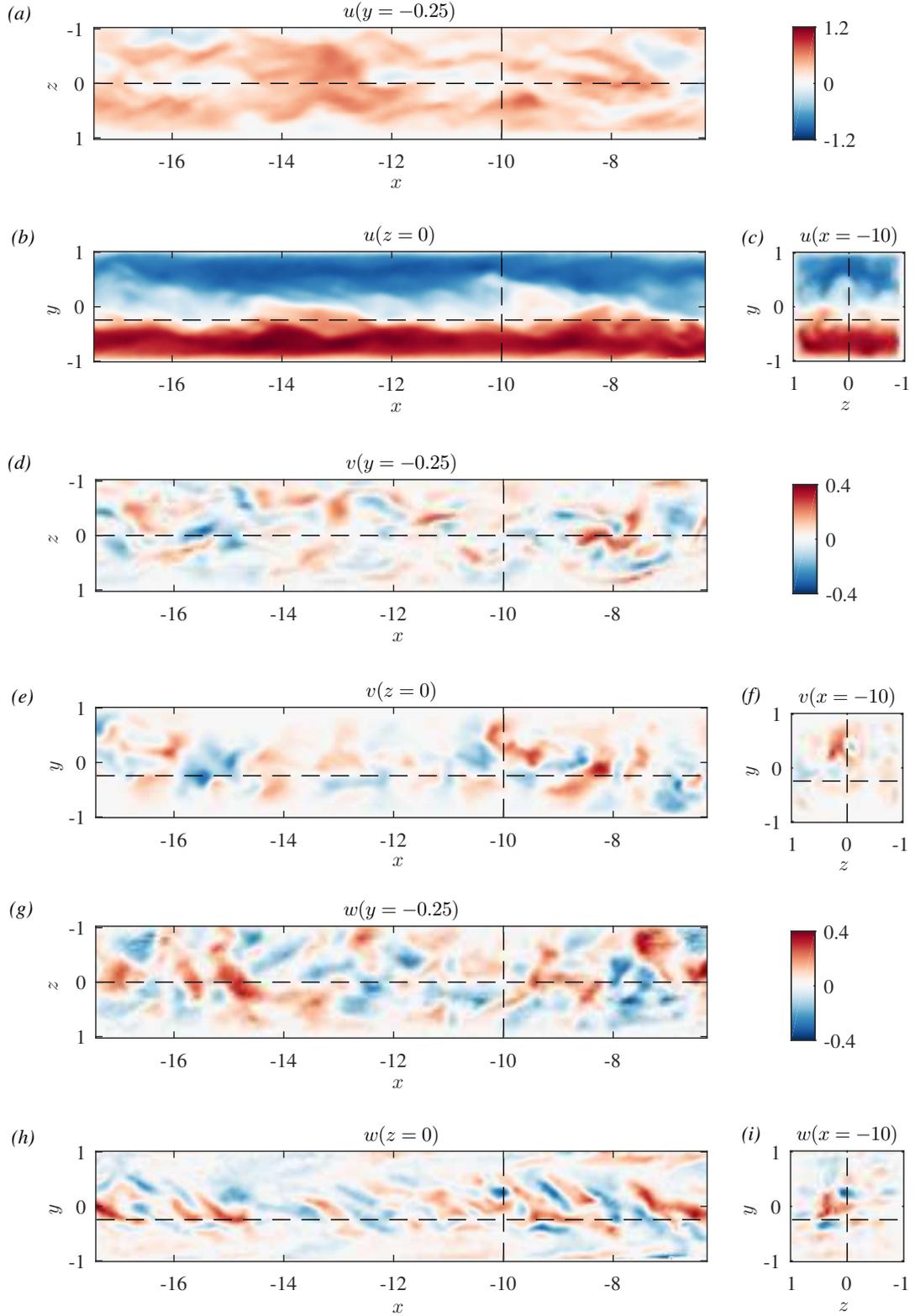}
\caption{{Instantaneous snapshots of all of the velocity components $u$ (a)-(c), $v$ (d)-(f), and $w$ (g)-(i) with no filtering of the data. On (a), (d), (g) the $y=-0.25$ horizontal plane; (b), (e), (h) streamwise vertical $z=0$ plane; and (c), (f), (i) the vertical $x=-10$ plane. The locations of the various slices are indicated by the dashed lines in (a)-(i)}. Note the different scale as typically $|u|>|v|$, $|w|$}\label{fig:instant}
\end{figure*}

\begin{figure*}[h!]
\centering
\includegraphics[width = 0.95\textwidth]{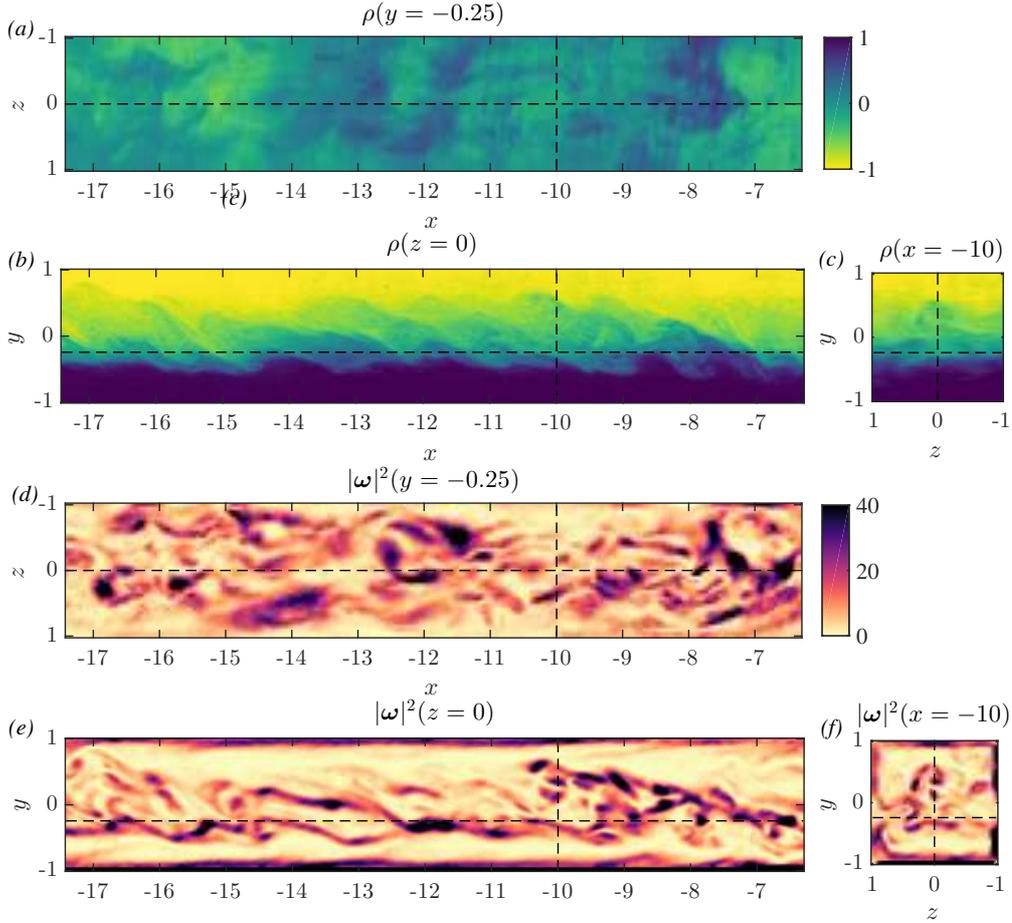}
\vspace{-2cm}
\caption{Instantaneous snapshots of density $\rho$ (a)-(c) and enstrophy $|${\boldmath${\omega}$}$|^2$ (d)-(f). On (a), (d) the $y=-0.25$ horizontal plane; (b), (e) streamwise vertical $z=0$ plane; and (c), (f) the vertical $x=-10$ plane. The locations of the various slices are indicated by the dashed lines in (a)-(f)}\label{fig:rho_vort}
\end{figure*}

\begin{figure*}[h]
\centering
\includegraphics[width = 0.95\textwidth]{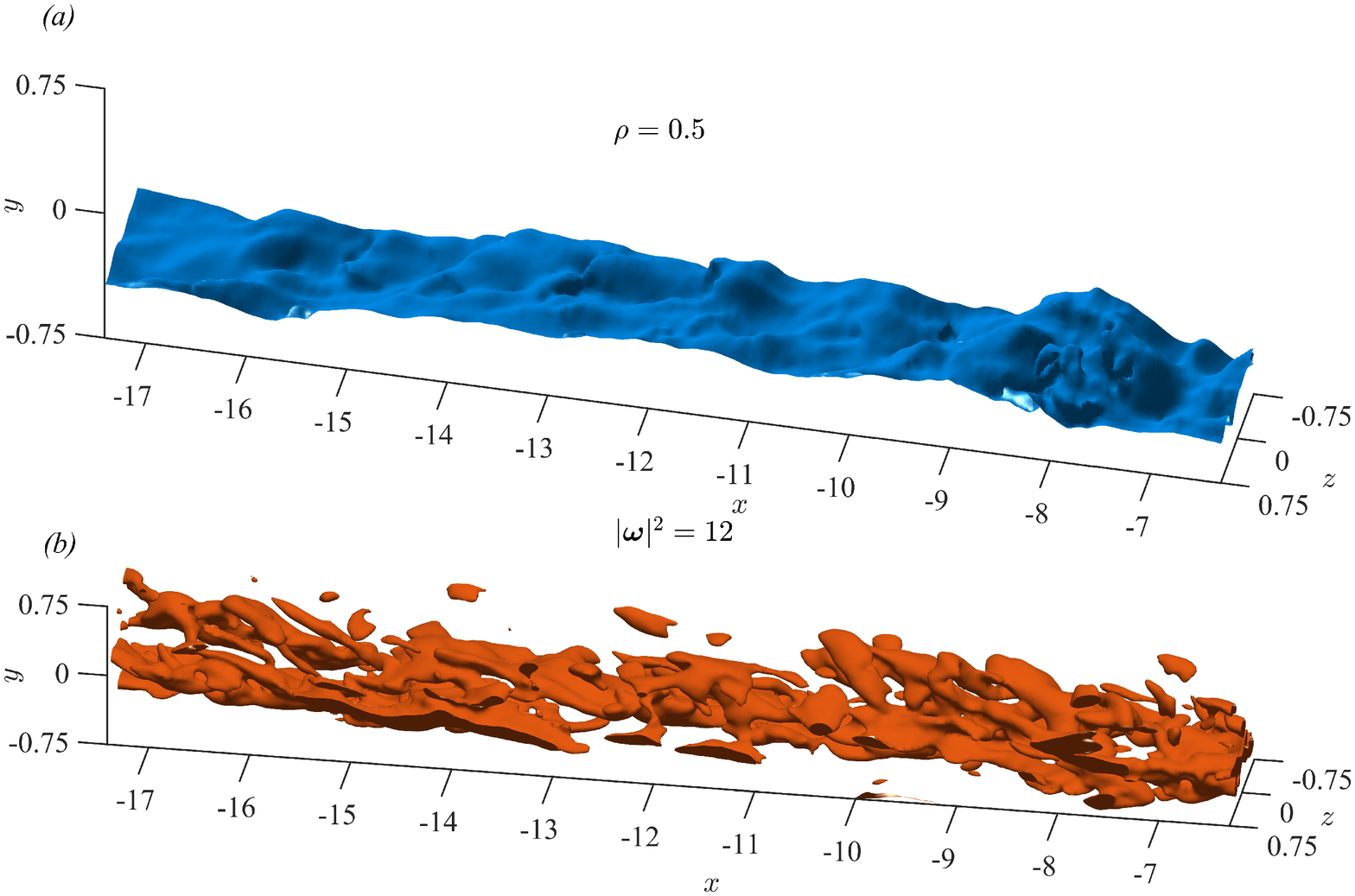}
\caption{{Instantaneous isosurfaces of (a) $\rho=0.5$ and (b) $|${\boldmath${\omega}$}$|^2=12$ in (b). For the data shown here,} an isotropic three-dimensional Gaussian filter with $\sigma_{filt} = 1$ vector spacing was used to smooth the data. For both plots, the vertical $y$ axis has been stretched by a factor of 2 to aid visualisation and the vertical and spanwise extents of the data have been limited to $[-0.75, 0.75]$ to avoid the signal at the boundaries obscuring the view}\label{fig:rho_vort_iso}
\end{figure*}

To get {an overall impression} of the flow, time-averaged quantities of $\mathbf{u}$ and $\rho$ are shown in figure \ref{fig:means}. In comparison to a laminar exchange flow, where there are two layers separated by a sharp density interface, mixing is evident in this relatively high $Re$ number flow. Here, thin {and} approximately well-mixed layers are confined to the horizontal boundaries at $y=\pm 1$. {These layers} are separated by a weakly-stratified interior (figure \ref{fig:means}(a) and (d)). Mean profiles of the streamwise velocity $u$ are also shown in figure \ref{fig:means}(b) and (h) and indicate the spanwise variation present within the flow due to the confining lateral boundaries at $z=\pm 1$. In the same geometry, the importance of this confinement to the flow at lower $Re$ has been investigated by \cite{lefauve2018}. Due to the confinement, they found a non-trivial modification to the classical Holmboe instability with significant 3D structure. The results of \cite{lefauve2018} demonstrate that, even at relatively low $Re$, 3D spatial measurements provide meaningful insight to the underlying dynamics that would be lost in classical planar measurements. 

Figure \ref{fig:means}(g) also shows that there is no significant spanwise variation of the density, and therefore no mean pressure gradients in the spanwise direction due to horizontal density gradients. Finally, indicative of the stability of the flow, the mean gradient Richardson number $\overline{Ri}_g(y,z)$ is also shown in figures \ref{fig:means}(c), (f), and (i). Note that here, 
\begin{equation}
\overline{Ri}_g(y,z) = \frac{1}{4}\frac{\langle N^2 \rangle_{x,t}}{{
\langle S^2 \rangle_{x,t}}}
\end{equation}
where 
\begin{equation}
\langle N^2 \rangle_{x,t} = \left(\frac{g'}{H}\right)\frac{\partial\langle \rho \rangle_{x,t}}{\partial y}
\end{equation}
is the buoyancy frequency associated with the averaged density profile $\langle \rho \rangle_{x,t}$ and 
\begin{equation}
\langle S^2\rangle_{x,t} = 4\left(\frac{g'}{H}\right)\left(\frac{\partial 
\langle u\rangle_{x,t}}{\partial y}\right)^2
\end{equation}
is the squared shear associated with the mean profiles of streamwise velocity $\langle u\rangle_{x,t}$, {with $\langle\cdot\rangle_{x,t}$ indicating the average in the streamwise direction $x$ and time.} From figures \ref{fig:means} (c) and (f) it is evident that there is an approximately constant region where $\overline{Ri}_g(y,z)\simeq 0.12$ that is confined vertically to the centre of the duct, associated with the weakly-stratified interior and approximately constant shear in this region (see figures \ref{fig:means} (d) and (e)). {This value of $\overline{Ri}_g(y,z)$ is expected given the turbulent nature of the flow, where values are typically $\overline{Ri}_g(y,z)<0.25$ \citep{holt1992numerical}.} 

Instantaneous data are shown in figures \ref{fig:instant} and \ref{fig:rho_vort}. Significant 3D structure is evident in all components of velocity $\mathbf{u}$ throughout the entire 3D domain, as shown in figure \ref{fig:instant}(a)-(h). At the same instant in time and on the same planes, the density field $\rho$ and the enstrophy $|${\boldmath$\omega$}$|^2$ are shown in figure \ref{fig:rho_vort}(a)-(e). These figures highlight the necessity of {time-resolved 3D fields} in analysing such flows, especially if coherent structures want to be extracted from the complex flowfield. Furthermore, the enstrophy could not be calculated without knowing gradient information of all components of $\mathbf{u}$. 

To demonstrate the {near-instantaneous nature} of the volumetric data, isosurfaces of {enstrophy} and density are shown in figure \ref{fig:rho_vort_iso}. To be able to capture structures, we only need to resolve the Eulerian timescales of the flow. This is distinct from needing to resolve the Lagrangian timescale between pairs of images to accurately determine the particles displacements (which are Lagrangian tracers) that would require a much faster frequency of acquisition over each volume. Crucially, quantities in the scan direction (including derivatives) are reliably resolved as the time between velocity/density subvolumes are only separated by $2\Delta t$ (therefore only double the $\Delta t$ needed to resolve the particle displacements) and the data are effectively skewed across the scan direction. The distorted nature of the data could be advected using the velocity information but this step has not been carried out on the data presented here. 

A further measure of the quality of the data can be inferred from $\zeta=|\mathbf{u}-\mathbf{u}_{div}|$ the $L_1$ norm of the difference between $\mathbf{u}$ and a divergence free field $\mathbf{u}_{div}$ calculated using the method described by \citet{wang2017weighted}. For the data shown here, the mean over the whole volume and all time is $\overline{\zeta}=2.5 \times 10^{-2}$ with standard deviation $\sigma=2.2\times10^{-2}$ (all in non-dimensional units) illustrating that the correction is small (note that the value of the mean corresponds to $\sim 0.17$ pixel/frame, the same order as our error estimate in \S\ref{measProc}). Given that all experimental data will have some non-zero divergence, and the fact that the measurements are near-instantaneous rather than {truly} instantaneous, we anticipate that $\zeta>0$. However, our measurements are validated as only a small numerical correction is required and we note that there is little qualitative difference between the $\mathbf{u}$ fields shown in this paper and the corresponding $\mathbf{u}_{div}$ fields.

\section{Discussion and Enhancements}\label{discuss}

There are a number of enhancements that extend the methodology presented in this paper. These enhancements make use of the same fundamental approach, i.e. superimposing a small translation of the light sheet independently of the traverse carriage motion, but extend it to `slower flows' (flows that are slow compared to the camera frame rate), flows with a strong out-of-plane motion, and experimental set-ups that require a larger depth of field.

\subsection{Slower Flows}\label{sec:modes}
{For flows that are slow relative to the camera frame rate, the methodology outlined so far would be slave to the time between light sheets. Slow flows require a larger $\Delta t$ so that the particle displacements are sufficient to obtain accurate velocity measurements. Therefore, for a given number of light sheets in a scan $2N$ the time taken to scan the volume will be $2N\Delta t$. However,} so far we have only {discussed} the simple `mode 0' operation of the system, where the mode number is given by $m$ in 
\begin{equation}
z_k = Z_{2k} = Z_{2(k+m)+1}.
\end{equation}
Here, $z_k$ defines the $z$ position where two light sheets overlap, with $Z_k$ the $z$ position of the $k$th laser pulse. Therefore, for mode 0, overlapping pairs of images each comprise of an even-numbered frame and its immediately following odd number, e.g. $z_0 = Z_{0} = Z_{1}$, $z_1 = Z_{2} = Z_{3}$, etc. However, for slower flows it is desirable to operate in higher modes where the images that are spatially coincident are separated in frame number. Operating in higher modes allows for greater temporal resolution for the complete scan that would otherwise be slave to the required $\Delta t$. Essentially, the effective $\Delta t$ used in the PIV calculation can be increased without having to decrease the scan rate or the number of subvolumes the volume is discretised into. However, operating in higher modes does make it necessary to displace the {laser} beam further and therefore requires bigger oscillating mirrors and larger aperture cylindrical lenses to accommodate the larger amplitude beam displacement.

\subsection{Mean Unidirectional Out-of-plane Motion}\label{meanMotion} 

{In traditional PIV methods (planar or stereo), strong out-of-plane motion can cause errors in the PIV algorithm (due to substantial loss of particles between frames) or demand a high acquisition rate (to minimise the loss of particles between frames) that increases the noise of the measurements due to the inevitable lower in-plane pixel displacement.} A potential use of the system, be it scanning or not, is the use of the mirrors on galvanometers to allow thin light sheets in configurations that would otherwise require a thick light sheet to accommodate the out-of-plane motion $w$. If it known beforehand that the flow to be measured has a unidirectional mean flow in the out-of-plane direction $\overline{w}$, the mirrors can be used to displace the light sheet in the $z$ direction a suitable amount to accommodate the motion, i.e. $D_m\sim\overline{w}\Delta t$. The anticipated out-of-plane motion could be determined by an initial experiment, with a thick light sheet, and then repeated with a thinner light sheet to improve the accuracy of the measurements.  

\subsection{Overcoming Depth of Field Limitations}\label{multTraverses}

{In some experimental set-ups it may be desirable to remove the depth of field limitation that currently limits the spatial extent of the scanning.} A possible extension of the system can be achieved by translating the cameras at the same time as the light sheet to remove the restraint on focal depth and keep the overlapping regions and magnification the same during the scanning. For liquid flows, this would ideally be a multi-traverse system. A two traverse system, one for the cameras and one for the optics producing the light sheet, has a distinct advantage in that the relative speeds of the camera traverse and the light sheet traverse can be tuned to keep the illuminated slice more closely in focus despite the change in depth of refractive index variations through the optical path, i.e. the relative distance of air and fluid between the cameras and the illuminated plane. Moreover, the optimal system would have four traverses in total, with a separate traverse for each camera. The four traverse system would have the added advantage that the overlapping region shared by all of the cameras could remain constant during the scanning. It is, however, not trivial to mount the cameras on a traverse such that the vibrations of the traverse do not impact the PIV measurements, which are highly sensitive to angular displacements. {For this reason, it could be better to have a single traverse for the cameras as it will allow for more widely spaced rails.} It is also conceivable to use computer-controlled lenses on all cameras to keep the light sheet in focus during the scan. However, this approach would only remove the focal depth constraint but there would still be variable magnification and non-constant overlapping regions during the scanning. Where possible, the best results, in terms of measurement accuracy, will be achieved by keeping the cameras static and making sure an adequate focal depth can be achieved.  

\subsection{Prospects}\label{fastCams}

For the current apparatus, the temporal resolution of the scanning is limited by the pulsed laser as, at a maximum, 100 pairs of discrete subvolumes can be illuminated yielding 100 2D3C velocity and 2D density slices. This restriction could be removed by using a pulsed laser system with a higher repetition rate at which point the limiting factor of the system would the cameras, the traverse system, or the galvanometers. Cameras with higher acquisition rates would be required along with a faster traverse system, as ultimately the time to scan the volume would be preferentially reduced instead of increasing the number of subvolumes acquired (due to the finite thickness of the light sheet). Moreover, for faster repetition rate lasers, the system will be restricted by the beam displacement possible by the mirrors on galvanometers. Ultimately, the $\Delta t$ required to adequately resolve the flow field is going to be independent of the repetition rate of the laser. Therefore, when sampling the flow field with a higher repetition rate laser and faster camera, to achieve a given $\Delta t$ between image pairs one could separate the images by $m$ frames where $m\simeq\Delta tf_{laser}$ (see \S\ref{sec:modes}). However, this approach means that the oscillating mirrors are going to have to displace the beam further to pair non-sequential light sheets. In theory this can be achieved by sufficiently spacing the two oscillating mirrors. In practice, the beam path should be made as small as possible from the laser source to the light-sheet-producing optics as the beam is inevitably diverging, reducing the light sheet quality. As well as this, the relative accuracy of the galvanometer rotation decreases as the spacing between the mirrors is increased.

The resolution of the in-plane measurements ($x$, $y$) is largely determined by the resolution of the cameras used in the imaging. To date, recording directly to the computer at 200 fps can be achieved with a $\sim$12 megapixel camera giving $\sim 10^5$ velocity vectors per image pair (assuming spacing of $8 \times 8$ pixels) and $\sim 10^7$ density measurements per image. Of course, the stereo PIV measurements can only be calculated where the two images of the PIV cameras overlap that, without translating the cameras during the measurements, puts a limit on the scanning direction $z$. Furthermore, for stationary cameras there is a limit in the scanning direction $z$ to keep adequate focusing across the whole volume (see \S\ref{multTraverses}). Tests suggest that the limit in the scanning direction is approximately half the field of view of the cameras (with a reduced aperture for the PLIF camera).

\section{Conclusions}\label{conc}

We have presented a novel addition to the scanning PIV/PLIF technique. Our new technique allows for fast, repeated scans of a measurement volume. The volume is assembled from a series of subvolumes that allow for the simultaneous measurement of all three-components of the velocity field and the density field. 

The new technique has been used to perform measurements on a stratified shear flow similar to the flow investigated by \cite{meyer2014stratified}. The volumetric measurements reveal a striking three-dimensional structure to the flow, both in single volumetric snapshots as well as time-averaged quantities.

The novel addition to the setup is the use of two mirrors on galvanometers to position the light sheet during the scanning. These mirrors allow thin, overlapping light sheets between pairs of images despite the light-sheet-producing optics being continuously translated by a traverse system to scan a large volume. 

For the current set-up, the volume is scanned in less than a second, limited by the repetition rate of the laser and the desired number of frames in the scanning direction, yielding a total of $O(10^6)$ velocity vectors and $O(10^7)$ simultaneous density measurements in each scan. The scanning period could be reduced, and the total yield of vectors/density measurements increased, with slightly different hardware using the same fundamental strategy described in this paper.

Ultimately, over a volume spanning up to approximately $80 \times 80 \times 50$ cm ($x$,$y$,$z$), a resolution of $500 \times 375 \times 100$ velocity vectors and $4000 \times 3000 \times 100$ simultaneous density measurements can be obtained in 1 s using the methodology outlined in this paper. Note that the two planes of information used to calculate the velocity field give independent density information through PLIF but the spatial resolution is still the same as the two planes are coincident spatially. In other words, in a volume of $O(10^5)$ cm$^3$ a total yield of vectors possible in 1 s is $O(10^7)$ along with $O(10^9)$ simultaneous density measurements. 

\section*{Acknowledgements}
We gratefully acknowledge the skills and expertise provided by the technical staff of the G.\ K.\ Batchelor Laboratory and Dr Mark Hallworth. Finally, we wish to thank the authors of \citet{wang2017weighted} who graciously gave us their code for the velocity divergence correction.

\bibliographystyle{agsm}
\bibliography{ScanningPaper_arXiv}

\end{document}